\documentclass[useAMS,usenatbib]{mnras}
\usepackage{graphicx,verbatim}
\usepackage[normalem]{ulem}
\usepackage{color,ulem}
\usepackage{cancel}
\usepackage{newtxtext}
\usepackage[T1]{fontenc} 
\usepackage{amsmath,amssymb}

\usepackage{caption}
\usepackage{subcaption}
\usepackage{cprotect}

\usepackage{lmodern}

\graphicspath{{./}{fig_new/}}

\newcommand{\eref}{equation~\,\ref}
\newcommand{\fref}{Fig.~\,\ref}
\newcommand{\aref}{Appendix~\,\ref}

\def \be{\begin{equation}}
\def \ee{\end{equation}}
\newcommand       \ba           {\begin{eqnarray}}
\newcommand       \ea           {\end{eqnarray}}
\def \la{\langle}
\def \ra{\rangle}
\def \bea{\begin{eqnarray}}
\def \eea{\end{eqnarray}}

\newcommand{\comments}[1]{}

\definecolor{webgreen}{rgb}{0,.5,0}
\definecolor{webbrown}{rgb}{.6,0,0}
\hypersetup{%
   colorlinks=true,hyperfootnotes=false,%
   breaklinks=true,%
   plainpages=false, bookmarksnumbered, bookmarksopen=true,
   bookmarksopenlevel=1,%
   urlcolor=webbrown, linkcolor=webbrown, citecolor=webgreen,
   }

\long\def\/*#1*/{}

\title[MRI driven dynamo]{Shedding light on the MRI driven dynamo in a stratified shearing box}   

\author[ Dhang et al.]
{Prasun Dhang$^{1,2}$,\thanks{E-mail:prasundhang@gmail.com}
{Abhijit B. Bendre$^{3}$,}
{Kandaswamy Subramanian$^{2,4}$}\\
$^{1}$JILA, University of Colorado and National Institute of Standards and Technology, 440 UCB, Boulder, CO 80309-0440, USA \\ 
$^{2}$IUCAA, Post Bag 4, Ganeshkhind, Pune 411007, India \\
$^{3}$Institute of Physics, Laboratory of Astrophysics, \'Ecole Polytechnique F\'ed\'erale de Lausanne (EPFL), 1290 Sauverny, Switzerland\\
$^{4}$Department of Physics, Ashoka University, Rajiv Gandhi Education City, Rai, Sonipat 131029, Haryana, INDIA
}
\begin{document}
\maketitle
\label{firstpage}
\begin{abstract}
We study the magneto-rotational instability (MRI) dynamo in a 
geometrically thin disc ($H/R\ll 1$) using
stratified zero net (vertical) flux (ZNF) shearing box 
simulations. We find that mean fields  and electromotive forces (EMFs) 
oscillate with a primary frequency $f_{\rm dyn} = 0.017$ $\Omega$ 
($\approx 9$ orbital period), but also  have
higher harmonics at $3f_{\rm dyn}$. 
Correspondingly, the current helicity,  
has two frequencies $2f_{\rm dyn}$ and $4f_{\rm dyn}$ respectively, which 
appear 
to be the beat frequencies
of mean fields and EMFs as expected from the magnetic helicity density evolution equation. 
Further, we 
adopt a novel inversion  algorithm called the  `Iterative Removal Of Sources' (IROS), 
to extract the turbulent dynamo coefficients in the mean-field closure 
using the mean magnetic fields and EMFs obtained from the shearing box simulation.
We show that  an $\alpha-$effect ($\alpha_{yy}$) is predominantly responsible for
the creation of the poloidal field from the toroidal field, while shear generates back a toroidal field from the poloidal
field; indicating that an $\alpha-\Omega$-type dynamo is
operative in MRI-driven accretion discs. We also find that both  strong outflow  ($\bar{v}_z$) and turbulent pumping ($\gamma_z$ ) 
transport  
mean fields away from the mid-plane.
Instead of turbulent diffusivity, they are the principal sink terms in the mean magnetic
energy evolution equation. We find encouraging evidence that 
a generative helicity flux is responsible for the  
effective $\alpha$-effect. Finally, we point out  potential limitations 
of horizontal ($x-y$) averaging  in defining the `mean' on the extraction of 
dynamo coefficients and their physical interpretations.
\end{abstract}


\section{Introduction}
\label{sect:intro}
The problem of angular momentum transport is a key concept in a rotationally supported accretion disc (for a review, see \citealt{Balbus_hawley1998}). The current consensus  
is that a weak magnetic field instability, namely magneto-rotational instability (MRI; \citealt{Velikhov1959, Chandrasekhar1960, Balbus_hawley1991, Balbus_hawley1992}) is responsible for outward angular momentum transport and drives mass accretion in a sufficiently ionized accretion disc (e.g. as in X-ray binaries, inner part of AGN discs). Although linear MRI ensures outward angular momentum transport, it must be studied in the non-linear phase to account for different observable phenomena in accretion discs.

MRI in an accretion disc is either studied in a local set-up (shearing box; \citealt{Balbus_hawley1992, Brandenburg1995, Hawley1995, Davis2010, Shi2010, Bodo2014,PFB16}) or in a global simulation (\citealt{Stone1999, Hawley2001, Hawley2013, Beckwith2011, Parkin2013, Hogg2016, Dhang2019, Dhang2023}). While a global approach is more desirable, it is computationally expensive. On the other hand, the shearing box approach offers an alternate path which is computationally less costly and can 
provide deep insights into the local processes in MRI-driven turbulence.

In the shearing-box approach (\citealt{Goldreich1965}), we expand fluid equations to the lowest order of   $H/R$, where $H$ is the density scale height and $R$ is the local radius. Therefore, this approach is valid only for geometrically thin discs with $H/R \ll 1$. Depending on whether the vertical component of gravity ($g_z=-\Omega^2 z$) (producing a vertically stratified gas density) is considered in the momentum equation or not, shearing box simulations are of two types: stratified ($g_z \neq 0$) and unstratified ($g_z=0$). Further, depending on whether the 
computational domain can contain net  vertical magnetic flux,  shearing box models can be classified into zero net flux (ZNF) and net flux (NF) models. Therefore, four possible combinations of the shearing-box model are: i) unstratified ZNF, ii) unstratified NF, iii) stratified ZNF and iv) stratified NF.  This work considers a stratified ZNF shearing box model to explore the MRI dynamo in saturation. 

Shearing box simulations provide a wide range of behaviour (e.g., convergence, turbulence characteristics etc.) depending on the shearing box model used (for details, we refer to readers to see Table 1 in \cite{Ryan2017}). However, it is to be noted that we will restrict our discussion to the isothermal (i.e. sound speed is constant) models where there is no explicit dissipation and the numerical algorithms provide the dissipation through truncation error
at the grid scale. In the presence of an NF, unstratified 
shearing box simulations show a convergence (in terms of accretion stresses) and sustained turbulence (\citealt{Hawley1995, Guan2009, Simon2009}). On the other hand, stratified NF simulations present different accretion stresses depending on the net flux strength and sustained turbulence (\citealt{Guan2011, Bai2013}). Unstratified ZNF models showed intriguing behaviour. Earlier  isothermal 
unstratified ZNF studies (\citealt{Fromang2007_I, Pessah2007}) found decreased accretion stress and turbulence with increased resolution, implying non-convergence.  However, later \cite{Shi2016} recovered convergence using a box with a larger vertical extent than the radial extent.  In contrast, earlier stratified ZNF models (\citealt{Davis2010}) suggested that the models are converged till the 
resolution $128/H$; however, recent studies (\citealt{Bodo2011, Ryan2017}) found the model loses convergent properties at higher resolution. 

The convergence problem is closely related to the magnetic energy
generation process in the MRI-driven flow. For the ZNF (absence of net flux) models, an MRI-driven dynamo must act  to overcome  the
diffusion and sustain the zero net flux in the accretion flow. Earlier  ZNF simulations in unstratified (\citealt{Shi2016}) and stratified (\citealt{Davis2010, Bodo2014, Ryan2017}) shearing boxes found MRI turbulence can self-generate large-scale magnetic fields attaining quasi-stationarity and sustaining turbulence. \cite{Riols2013} suggested that the non-linear MRI does not behave 
like a linear instability; rather, it provides a pathway for saturation via a subcritical dynamo process. This leads to the question of what kind of dynamo  can be sustained in the MRI-driven accretion flow, small-scale or large-scale?  The lack of convergence in ZNF models was attributed  to the low numerical Prandtl number (\citealt{Fromang2007_I}; however, see \citealt{Simon2009}) and hence the inefficiency of
small-scale dynamo to operate at small Prandtl number (\citealt{Schekochihin2005, Bodo2011}). 
However, it is unclear what happens when convergence is recovered in unstratified ZNF simulations with tall boxes (\citealt{Shi2016}).

Studying MRI dynamo is also important for understanding the generation 
of coherent large-scale magnetic fields determining the 
level of transport (\citealt{Johansen2009, Bai2013}) and outflows from the accretion disc (\citealt{Rekowski2003,Stepanovs2014, Mattia2022}). MRI, in principle, can generate  magnetic fields coherent over several scale-heights (\citealt{Dhang2023}) and acts locally as a mean field in the absence of any external flux influencing convergence and  the disc dynamics.

Generally, stratified models generate a more coherent large-scale field over the unstratified models (for a comparison, see \citealt{Shi2016}). 
Cyclic behaviour of azimuthally averaged magnetic fields (mean fields), popularly known as the butterfly diagram, is a typical 
feature observed in the stratified shearing box simulations (\citealt{Brandenburg1995, Gressel2010, Bodo2014, Ryan2017, Gressel2022}). 
However, note that the presence of a strong magnetic net flux (\citealt{Bai2013, Salvesen2016}), convection (\citealt{Hiorse2014, Coleman2017}) etc. can alter the periodicity in the butterfly diagram. Although the cyclic behaviour of mean 
fields can be explained by invoking the interplay between shear and helicity (\citealt{Brandenburg1997,Gressel2015}), some features, such as upward migration of the mean fields, still demand an explanation.

Several studies attempted to understand the underlying mechanisms of 
MRI dynamo using different approaches. While some of the studies (\citealt{Lesur2008, Bai2013, Shi2016, Begelman2023}) invoked toy models to complete the generation cycles of radial and azimuthal fields, others
(local: \citealt{Brandenburg_Radler2008, Gressel2010, Shi2016, Gressel2022, Mondal2023}, global: \citealt{Dhang2020}) used mean-field theory to investigate the 
large-scale field generation in the MRI-driven turbulent accretion flow. Most of the studies characterising the turbulent 
dynamo coefficients in the regime of mean-field dynamo theory used state of the art `Test Field' (TF) method (\citealt{Gressel2010, Gressel2015}), while a few used direct methods such as linear regression (\citealt{Shi2016}), singular value decomposition (SVD; \citealt{Dhang2020}) to calculate dynamo coefficients in post-process or statistical simulations to carry out combined study of  the large-scale dynamo and angular-momentum transport in accretion discs (\citealt{Mondal2023}).   In this work, we use a direct method, a variant of the cleaning algorithm ( H\"ogbom CLEAN method; \citealt{Hogbom1974}), called `Iterative Removal Of Sources' (IROS; \citealt{Hammersley1992}); mainly used in 
astronomical  
image construction to analyse MRI-dynamo in the mean-field dynamo paradigm. We modified the IROS method according to our convenience (for details, see section \ref{sect:IROS}, also  see \citealt{Bendre+2023}) and used it to determine the dynamo coefficients by post-processing the data obtained from the stratified ZNF shearing box simulation. 

The paper is organised as follows. In section \ref{sect:method}, we describe details of shearing box simulations, basics of mean field closure used and techniques of the IROS method. 
Section \ref{sect:result} describes the evolution of MRI to a non-linear saturated 
state, spatio-temporal variations of mean magnetic fields, EMFs and periodicities present in different observables.  The spatial profiles of calculated turbulent 
dynamo coefficients, the reliability of the calculation method (using both EMF reconstruction and a 
1D dynamo model)
and contributions of each coefficient to the mean magnetic energy equation are described in  section \ref{sect:IROS}. In section \ref{sect:discuss}, we discuss the plausible reasons behind 
different periodicities present (in mean magnetic fields, EMFs and helicties), comparison of our work with the previous works, 
the possible importance of a generative helicity flux 
and limitations of the averaging scheme and mean-field closure used in decoupling contributions from different dynamo coefficients. Finally we summarized our key results in section \ref{sect:summary}.

\section{Method}
\label{sect:method}
The current work involves performing shearing box simulations of MRI-driven accretion flow, along with extracting dynamo coefficients using the mean-field dynamo model. In this section, we discuss details of the shearing-box simulation set-up, an introduction to the mean-field dynamo model and the IROS method used to determine turbulent dynamo coefficients using the simulated data.
\subsection{Shearing-box simulation}
\label{sect:shearingbox}
We perform stratified zero net flux (ZNF) shearing box simulations to study the MRI driven dynamo in a geometrically thin disc ($H/R \ll 1$). To do that, we solve ideal MHD equations
in a Keplerian shearing box  given by
\bea 
&& \frac{\partial \rho }{\partial t} + \nabla \cdot \left( \rho \mathbf{v} \right) = 0,\\
&& \frac{\partial \rho \mathbf{v}}{\partial t} + \nabla \cdot \left(\rho \mathbf{v}\mathbf{v} - \mathbf{B} \mathbf{B} \right) + \nabla P = \rho \mathbf{g}_s - 2 \Omega \hat{z} \times \rho \mathbf{v},\\
&& \frac{\partial \mathbf{B}}{\partial t} = \nabla \times \left(\mathbf{v} \times \mathbf{B} \right)
\eea 
using the {\tt PLUTO} code (\citealt{Mignone2007}) with $x, \ y, \ z$ as the radial, azimuthal and vertical directions respectively. Here, $ \rho, \ P, \ \mathbf{v}$ and $ \mathbf{B}$ denote density, thermal pressure, velocity and magnetic fields, respectively. The terms $\mathbf{g}_s = \Omega^2 \left(2 q x \hat{x} - z \hat{z}\right)$ and  $2 \Omega \hat{z} \times \rho \mathbf{v}$ represent the tidal expansion of the effective gravity and the Coriolis force respectively with $\Omega$ denoting orbital frequency. We use
an isothermal equation of state 
\be 
P=\rho c^2_{s}.
\ee 
Therefore, we do not need to solve the energy equation. Additionally, we use constrained transport (\citealt{Gardiner_stone2005}) to maintain divergence free condition 
\be
\nabla \cdot \mathbf{B}=0
\ee 
for magnetic fields. We use the HLLD solver
(\citealt{Miyoshi2005}) with second-order slope-limited reconstruction. Second-order Runge–Kutta (RK2) is used for time integration with the CFL number 0.33. Also note that despite our shearing-box model lacking explicit dissipation, we refer to it as the direct numerical simulation (DNS).

We initialize an unmagnetized equilibrium solution with density and velocity given by
\bea 
&& \rho = \rho_0 \ {\rm exp}\left(-{\frac{z^2}{2H^2}} \right), \\
&& \mathbf{v} = -q \ \Omega \ x \ \hat{y}
\eea  
where $q=1.5$ and  $\rho_0$ is the mid-plane ($z=0$) density and 
\be 
H = \frac{c_s}{\Omega} 
\ee 
is the thermal scale height. We set $\rho_0=c_s=\Omega=1$, so that $H=1$. Unless stated otherwise, all the length and time scales are expressed in units of $H$ and $\Omega^{-1}$ respectively. We initialize a ZNF magnetic field given by
\be 
\mathbf {B} = \sqrt{\frac{2}{\beta_0}} \ \sin \left( \frac{2 \pi x}{L_x} \right) \ \hat{z}
\ee 
with $\beta_0=10^4$ defining the strength of the field and  $L_x, \ L_y, \ L_z$ denoting the size of the shearing-box.

Our computational domain extends from $-L_x/2 < x < L_x/2$, $-L_y/2 < y < L_y/2$ and $-L_z/2 < z < L_z/2$. It has been found in earlier studies  
that shearing box results depend on the domain size; larger boxes tend to capture dynamo better than their smaller counterparts as well as smaller boxes demonstrate a transition to anomalous behaviour (e.g. see \citealt{Simon2012, Shi2016}). To avoid these
discrepancies, we choose a shearing box of size $L_x \times L_y \times L_z=3H \times 12H \times 8H$ with a grid resolution $N_x \times N_y \times N_z = 96 \times 192 \times 256$ giving rise to a resolution of $32/H$ in the vertical direction. However, we must admit that there exists an issue with the convergence in stratified ZNF models as discussed in the section \ref{sect:intro}. We reserve the dependence of MRI dynamo on numerical resolution as a topic of  future research investigation.

We use periodic and shearing-periodic (\citealt{Hawley1995}) boundary conditions in the $y$ and $x-$ boundaries, respectively. Outflow boundary conditions are implemented in the vertical ($z$) boundaries. A gradient-free condition is maintained for scalars and tangential components of vector fields at the boundaries. At the same time,  $v_z \ge  0$ for $z>0$ and $v_z \le  0$ for $z<0$ 
is set to  restrict mass inflow into the domain at vertical boundaries. The z-component of the magnetic field is set by the divergence-free condition of the magnetic field.

Turbulent dynamo coefficient estimation involves analysis of time series of mean magnetic fields and EMFs obtained in shearing box simulation. Therefore, we dump the data quite frequently with data dumping interval $\Delta t=0.2$ $\Omega^{-1}$ and run it till $t=300$ $\Omega^{-1}$ to have enough number data points in the time series of mean magnetic fields and EMFs.

\subsection{Mean field closure}
Before describing the details of mean field dynamo theory and the closure used, we define what is meant by `mean' and `fluctuation' in our work.  
We define mean magnetic fields ($\mathbf B$) as the $x-y$-averaged values as follows
\be
\label{eq:mean_def}
\bar{\mathbf B}(z,t) = \frac{1}{L_x L_y} \int_{-L_x/2}^{L_x/2}\int_{-L_y/2}^{L_y/2} \mathbf B(x,y,z,t) \ dx \ dy.
\ee
Fluctuating magnetic fields are defined as 
\be
\label{eq:fluc_def}
{\mathbf B}^{\prime}(x,y,z,t) = {\mathbf B}(x,y,z,t) - \bar{\mathbf B}(z,t).
\ee
Mean and fluctuations of the $x-$ and $z-$ components of the velocity are defined in the same way as those for magnetic fields, while the mean and fluctuation of $y-$component of velocity are defined as  
\be
\bar{v}_y  = -q \Omega x, \ \  v^{\prime}_y = v_y - \bar{v}_y.
\ee
If we decompose the magnetic and velocity fields into mean and fluctuation and insert them into the magnetic field induction equation, we obtain the mean-field equation
\be
\label{eq:mean_field_eq}
\frac{\partial \bar{\mathbf B}}{\partial t} = \nabla \times \left(\bar{\mathbf v} \times \bar{\mathbf B} \right) + \nabla \times \bar{\mathcal{E}};
\ee 
where we assume that microscopic diffusivity is vanishingly small (ideal MHD limit).
Here mean EMF
\be
\bar{\mathcal{E}} = \overline{v^{\prime} \times B^{\prime}}
\ee
appears as a source term in  equation \ref{eq:mean_field_eq}.
The crux of the mean-field dynamo theory is how to express mean EMF in terms of the mean magnetic fields.
In general, the usual mean-field closure (\citealt{Radler1980, Brandenburg2005, anv_kan_book}) is given by, 
 \be
 \label{eq:basic_closure}
 \bar{\mathcal{E}}_i (z) = \alpha_{ij}(z) \  \bar{B}_j (z) - \eta_{ij}(z) \  \bar{J}_j,
 \ee
 where we neglect higher than  the first-order spatial  derivatives  and time derivatives of mean magnetic fields and $\alpha_{ij}$, $\eta_{ij}$ are the turbulent dynamo coefficients 
 which characterize the dynamo; and $\bar{J}_j=\epsilon_{jzl} \partial_z \bar{B}_l (z)$ is the current. Further, while calculating turbulent dynamo coefficients using direct methods (e.g. SVD (\citealt{Bendre2020, Dhang2020}), linear regression (\citealt{Squire2016, Shi2016}), it is also assumed that $\alpha_{ij}$, $\eta_{ij}$ are constant in time. However, we find that in our simulation of MRI-driven accretion flow, the current helicity, which is potentially
 a primary component determining the $\alpha_{ij}$, shows a reasonably  
 periodic change over time with a period half the dynamo-period (for details, we refer the reader to  section \ref{sect:kin_curr_helicity}). This time-dependent feature of current helicity leads to considering a heuristic  mean field closure defined as
 \be
 \label{eq:closure}
 \bar{\mathcal{E}}_i (z) = \left( \alpha_{ij}^0 + \alpha^{1}_{ij} \cos(2\Omega_{\rm dyn}t+\phi) \right) \bar{B}_j (z) - \eta_{ij} \bar{J}_j 
 \ee
 to capture the time dependence in $\alpha_{ij}$. Here $\alpha^{0}_{ij}$ and $\alpha^{1}_{ij}$ are the time-independent (i.e. DC component) and time-dependent parts of $\alpha_{ij}$ respectively and $\Omega_{\rm dyn} = 2 \pi f_{\rm dyn} =2 \pi/T_{\rm dyn}$, with $f_{\rm dyn}$ and $T_{\rm dyn}$ are the dynamo frequency and  period respectively. 
 Further, one expects  that $\eta_{ij}$ to be dominated by a DC component, because $\eta_{ij}$-s are generally
 determined by the turbulent intensity of the flow, not by helicities.
 Thus for simplicity, we adopt a time-independent $\eta_{ij}$.

\subsection{Dynamo coefficient extraction method - IROS}
\label{sect:IROS}
We solve \eref{eq:closure} in a least-square sense to extract the turbulent dynamo coefficients 
($\alpha_{ij}^0$, $\alpha_{ij}^1$ and $\eta_{ij}$) using  mean magnetic fields $\bar{B}_{i}(z,t)$ and EMFs $\bar{\mathcal{E}}_{i} (z,t)$ (with $i \in \{x,y\}$)  obtained from shearing-box simulations described in section \ref{sect:shearingbox}. Further we assume that these dynamo
coefficients stay statistically unchanged during the 
quasi-stationary phase of evolution, i.e. the coefficients are independent of time. Hence, all the dynamo coefficients are only dependent on the vertical coordinate $z$.

As a first step of coefficient determination in this under-determined system, we construct the time series of
length $N$, of mean EMFs $\bar{\mathcal{E}}_i(z,t_1\hdots t_N)$, mean magnetic fields $\bar{B}_i(z,t_1\hdots t_N)$ and currents
$\bar{J}_i(z,t_1\hdots t_N)$ from the DNS ($i\in\{x,y\}$). With these time series, we rewrite 
\eref{eq:closure} at any particular $z=z'$ as 

\begin{align}
 \mathbf{Y}(z',t) = \mathbf{A}(z',t)\,\mathbf{X}(z'), 
\label{eq:linear}    
\end{align}
where the matrices $\mathbf{y}$, $\mathbf{A}$ and $\mathbf{x}$ are defined as, 
\begin{align}
\mathbf{Y}(z',t)  =&
\begin{bmatrix}
   \mathcal{E}_x(z',t_1) & \mathcal{E}_y(z',t_1)\\[4pt]
   \mathcal{E}_x(z',t_2) & \mathcal{E}_y(z',t_1)\\
   \vdots&\vdots\\
   \mathcal{E}_x(z',t_N) & \mathcal{E}_y(z',t_1)
\end{bmatrix}
\nonumber\\ \nonumber\\
\mathbf{A}^{\intercal}(z',t)=&
\begin{bmatrix}
\bar{B}_x(z',t_1)   & \bar{B}_x(z',t_2) &\hdots \bar{B}_x(z',t_N)\\[4pt]
\bar{B}_y(z',t_1)   & \bar{B}_y(z',t_2) &\hdots \bar{B}_y(z',t_N)\\[4pt]
\mathcal{C}_x(z',t_1)   & \mathcal{C}_x(z',t_2) &\hdots \mathcal{C}_x(z',t_N)\\[4pt]
\mathcal{C}_y(z',t_1)   & \mathcal{C}_y(z',t_2) &\hdots \mathcal{C}_y(z',t_N)\\[4pt]
-\bar{J}_x(z',t_1)  & -\bar{J}_x(z',t_2) &\hdots -\bar{J}_x(z',t_N)\\[4pt]
-\bar{J}_y(z',t_1)  & -\bar{J}_y(z',t_2) &\hdots -\bar{J}_y(z',t_N)
\end{bmatrix}\nonumber\\ \nonumber\\
 \mathbf{X}(z')  =&
\begin{bmatrix}
    \alpha_{xx}^0(z') &\alpha_{yx}^0(z')\\[4pt]
    \alpha_{xy}^0(z') &\alpha_{yx}^0(z')\\[4pt]
    \alpha_{xx}^1(z') &\alpha_{yx}^1(z')\\[4pt]
    \alpha_{xy}^1(z') &\alpha_{yx}^1(z')\\[4pt]
    \eta_{xx}(z')   &\eta_{yx}(z')\\[4pt]
    \eta_{xy}(z')   &\eta_{yy}(z')
\end{bmatrix}.
\end{align}
Here the terms $\mathcal{C}_i(z',t_l)=\bar{B}_i(z',t_l)\,
\cos{(2\Omega_{\rm dyn}t_l+\phi)}$ ($\forall i\in\{x,y\}$) 
which take into account the time dependent part of $\alpha_{ij}$. 
For simplicity, we assume $\phi$ to be zero.

Our task is to determine the dynamo coefficients ($\mathbf{x}$) 
by pseudo-inverting \eref{eq:linear}. This task is complicated 
firstly by the fact that both components of mean-field and 
current have additive correlated noise and secondly by the fact 
that the $y$ component of the mean-field is typically much 
stronger compared to the $x$ component, due to the rotational 
shear (and by consequence the $x $ component of current is
much stronger than its $y $ component). Typical schemes of
the least square minimisation in such cases tend to
underestimate the dynamo coefficients that are associated
with the $x$ component of mean-field (i.e. the 
coefficients $\alpha^0_{ix}$ and $\alpha^1_{ix}$), and those
with the $y$ component of mean current (i.e. the coefficients 
$\eta_{iy}$).
To circumvent these issues, we rely upon the 
IROS method (Iterative removal of sources) \citep{Hammersley1992} 
that we have recently adapted for such inversions in the dynamo
context \citep{Bendre+2023}. This method is based on H\"ogbom 
clean algorithm (\citealt{Hogbom1974}) used in Radio Astronomy 
to construct an image by convolving multiple beams, 
iteratively locating and subtracting out the strongest source 
to model the rest of the dirty image. It is particularly useful
when the relative contribution of some of the beams to the 
final image happens to be negligible. Such a situation is 
analogous to have only a few of the columns of $\mathbf{A}$ 
(the beams) largely contribute to $ \mathbf{y}$ (an image). 
A brief outline of the method is as follows. 

Firstly, at any particular $z=z' $ we set all the dynamo 
coefficients, $ \alpha^0_{ij}(z')$, $ \alpha^1_{ij}(z')$ 
and $\eta_{ij}(z')$ to zero, (i.e we set $\mathbf{X}(z')
=0$). Then to compute these coefficients, we 
iteratively estimate their magnitudes as follows.
To derive the zeroth order estimates of these coefficients 
we fit every $ i^{\rm th} $ column of $ \mathbf{Y}(z',t)$ 
 denoted as $\mathbf{Y}_i(z',t)$), 
against the individual columns of $ \mathbf{A}( z' ) $ 
(denoted as $ \mathbf{A}_k( z' ) $) separately as lines. 
Slopes and chi-square errors ($\chi^2_{ik}(z')$) 
associated with each fit are recorded. 
The individual chi-square errors are defined as $\chi^2_{ik}
(z')=\sum_i(\mathbf{Y}_i-\mathbf{A}_k \,\mathbf{X}_{ik})^2$).
Then the best fitted dynamo coefficient (the one 
which has the least chi-square error) is updated by 
adding to it, its zeroth order estimate multiplied 
by a small factor ($\epsilon<1$), called the loop-gain, 
while other coefficients are kept constant. For example, 
if the chi-squared error associated with the line fit 
$\mathcal{E}_x(z',t_1\hdots t_N)$ versus $\bar{B}_y(z',
t_1\hdots t_N)$ (i.e. $\chi^2_{12}(z')$) is the least 
and if the slope is $m$ then $ \mathbf{X}_{2,1}
(z')$ (i.e. $\alpha^0_{xy}$) is updated by adding
to it a factor of $m\epsilon$. Subsequently, the contribution to the EMF 
associated with the best fitted coefficient, also 
multiplied by the $\epsilon$ is subtracted from the 
corresponding EMF component. For instance, using 
the same example, a factor of $ \epsilon\, \alpha_{xy}
(z')\,\bar{B}_y(z',t_1\hdots t_N) $ is subtracted from 
$ \mathcal{E}_x( z',t_1\hdots t_N) $. This residual EMF 
is then used as an actual EMF component to further
compute higher order estimates of dynamo coefficients, 
and this process is repeated a suitable number of times 
until either all the dynamo coefficients converge to their 
respective constant values or all four chi-squared errors 
get smaller than a certain predefined threshold. All the 
aforementioned steps are then repeated at every $z=z'$.

We apply this method with $ \epsilon=0.1 $ for five hundred 
refinement loops to the time series of EMFs, mean-fields
and currents obtained from the DNS data. While constructing 
these time series (from $t=100 \Omega^{-1}$ to $300\Omega^{
-1}$) with data dumping interval $\Delta t_{\rm dump} =0.2\ 
\Omega^{-1}$ we make sure that they  correspond to the 
quasi-stationary phase of the magnetic field evolution. 

IROS method does not provide an estimate of errors on the 
calculated coefficients directly. We, therefore, calculate a 
statistical error of the dynamo coefficient by considering 
the five different realizations of time series. We construct 
five different time series of mean fields, currents and EMFs 
by skipping four data points in the time series. 
Specifically, the time series $(t_1,t_2,\hdots t_N)$ (of all 
components of mean-field, current and EMF) are split into 
$(t_1,t_6\hdots)$, $(t_2,t_7\hdots)$, $(t_3,t_8\hdots)$, 
$(t_4,t_9\hdots)$ and $(t_5,t_{10}\hdots)$. 
We use these time series to calculate five sets of dynamo 
coefficients and calculate their standard deviations to 
represent the errors on the calculated coefficients.

\section{Results: saturation of MRI, mean fields and EMF-s}
\label{sect:result}

We now turn to the results of our 
shearing box simulation of MRI in a geometrically thin disc, investigate its dynamo action in  addition to  discussing several 
important properties {which illuminates the nature of the MRI dynamo.
Most of our analysis of magnetic field generation will focus on the saturated state of MRI, 
when the disc is in the quasi-stationary phase.

\subsection{Saturation of MRI}
\begin{figure}
    \includegraphics[scale=0.55]{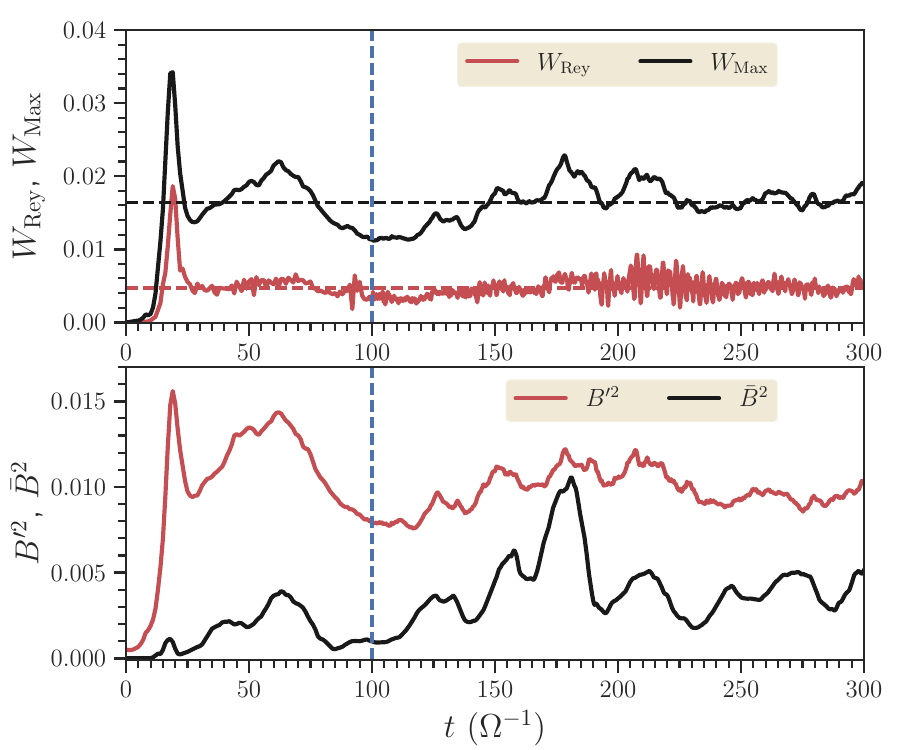}
 \caption{Top panel: Time history of Reynolds ($W_{\rm Rey}$) and Maxwell ($W_{\rm Max}$) stresses. Bottom panel: time history of the volume-averaged mean ($\bar{B}^2$) and fluctuating ($B^{\prime 2}$) magnetic energies.  }
   \label{fig:stress_b}
 \end{figure}

 First, consider  the time evolution of accretion stresses 
and magnetic energies. This will also allow us to determine  
the quasi-stationary phase of the MRI-driven turbulence.
The top panel of Fig. \ref{fig:stress_b} shows the time history of accretion stresses (Reynolds and Maxwell). Normalized Reynolds  and Maxwell stresses are defined as
\bea
&& W_{\rm Rey} = \frac{\la \overline{\rho v^{\prime}_x v^{\prime}_y} \ra_V }{\la p_{g} \ra_z}, \\
&& W_{\rm Max} = \frac{\la \bar{B}_x \bar{B}_y \ra_z + \la \overline{B^{\prime}_x B^{\prime}_y} \ra_z }{\la p_{g} \ra_z}, 
\eea 
where the averages are done over the whole volume. Here, $\bar{.}$ and  $\la . \ra_z$ indicate the average over $x-y$ and $z$ respectively. Reynolds stress is due to the correlation of fluctuating velocity fields, while Maxwell stress is composed of a correlation between the fluctuating components as well as that between the mean components of the magnetic fields. Both the stresses grow exponentially during the linear regime of MRI, and eventually saturate around an average value when MRI enters into the non-linear regime. In our simulation, we find the volume and time-averaged (within the interval $t=(100-300) \ \Omega^{-1}$) values of Reynolds and Maxwell stresses are to be $ W_{\rm Rey, av} = 0.0048$ and $ W_{\rm Max,av}  = 0.0167$ respectively.  The ratio of Maxwell to Reynolds stress is $  W_{\rm Max,av} / W_{\rm Rey,av} =3.5$, close to $4$, as predicted by \citealt{Pessah2006} for $q=1.5$ and similar to what is found in 
earlier numerical simulations (\citealt{Nauman2015, Gressel2015}).

The bottom panel of Fig. \ref{fig:stress_b} shows how the volume-averaged mean ($\la  \bar{B}^2 \ra_z $) and fluctuating ($ \la  \overline{B^{\prime 2}} \ra_z $) magnetic energies evolve over time. Like accretion stresses, magnetic energies oscillate about an average value in the quasi-stationary phase after the initial exponentially growing phase. It is also worth noting that the mean part of the magnetic field shows a larger time variation than the fluctuating part of the magnetic field. We point out an important point that the fluctuating magnetic field is stronger than the mean magnetic field, and the implication of this will be discussed in the latter part of the paper.

We see in Fig. \ref{fig:stress_b} that the accretion stresses and magnetic energies start saturating around $t=40 \ \Omega^{-1}$. However, to remain safer, we consider the simulation in the time range $t=(100-300) \ \Omega^{-1}$  for dynamo coefficient calculation in the quasi-stationary state.

 \subsection{Evolution of mean fields and EMFs}
 \label{sect:mean_field_emf_evo}
 \begin{figure*}
    \includegraphics[scale=0.58]{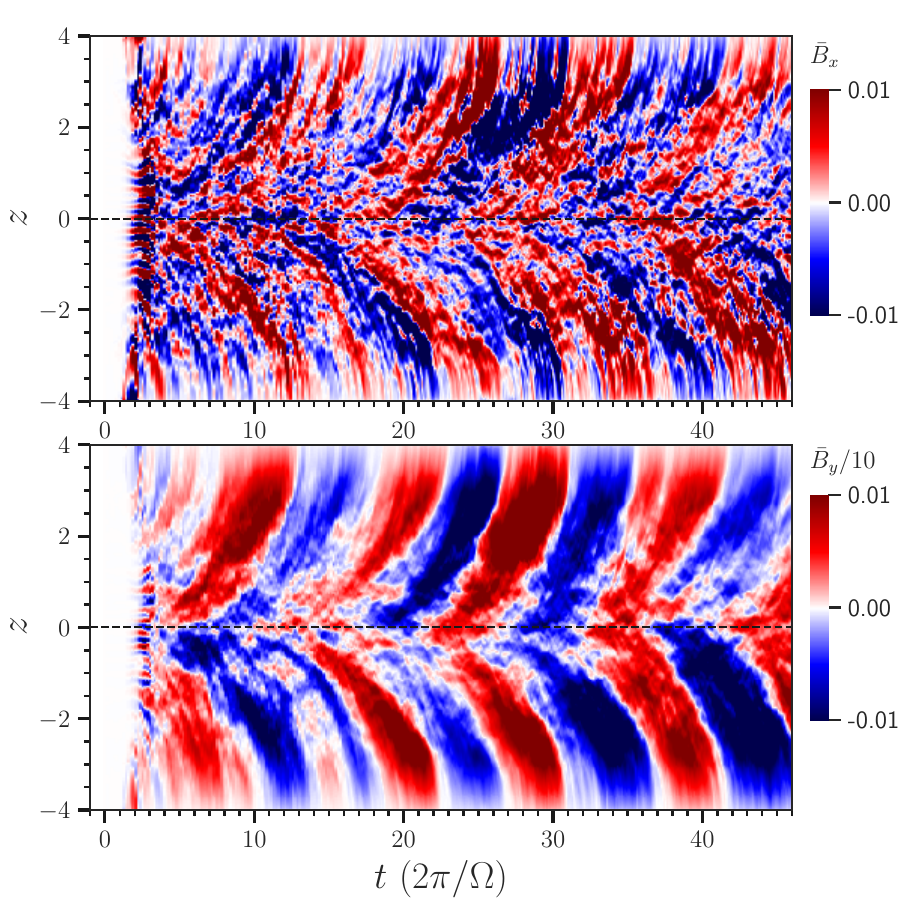}
     \includegraphics[scale=0.58]{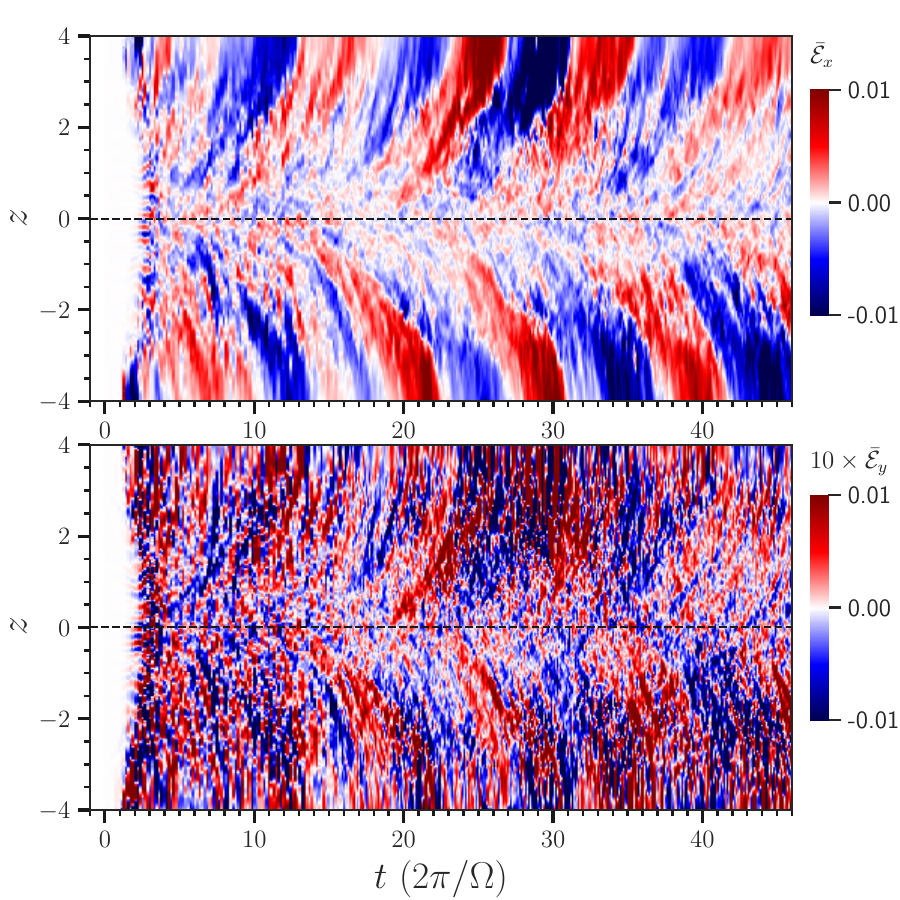}
 \caption{Spatio-temporal variation of mean magnetic fields, $\bar{B}_x$ (top left panel),  $\bar{B}_y$ (bottom left panel) and mean EMFs $\bar{\mathcal{E}}_x$ (top right panel) and $\bar{\mathcal{E}}_y$ (bottom right panel). Mean magnetic field component $\bar{B}_y$  and y-component of EMF $\bar{\mathcal{E}}_x$ show a coherent change in space and time (with a time period $\approx 9$ orbital period ($2\pi/\Omega$)), while the spatio-temporal patterns in $\bar{B}_x$ and $\bar{\mathcal{E}}_y$ are less coherent.}
   \label{fig:b_emf_mean}
 \end{figure*}
 
The most preliminary diagnostic of the dynamo is to look at the spatio-temporal variation of the mean magnetic fields, popularly known as the butterfly diagram (e.g. see the review by \citealt{Brandenburg2005}).
Fig. \ref{fig:b_emf_mean} shows the butterfly diagrams for mean magnetic fields $\bar{B}_x$ and $\bar{B}_y$ along with the mean EMFs $\bar{\mathcal{E}}_x$ and $\bar{\mathcal{E}}_y$. Here we note that the mean EMF 
acts as a source term in the 
mean magnetic field energy evolution equation. In particular, $\bar{\mathcal{E}}_y$} is responsible for the generation of poloidal field (here $\bar{B}_x$) from a toroidal one due to an $\alpha$-effect, which itself 
 naturally emerges by a combined action of stratification and rotation (\citealt{Krause_Raedler1980}) in our stratified shearing box simulation.   At an early stage of evolution (around $t\approx 2$ orbital period), both mean fields and EMFs show lateral stretches with changing the sign in the vertical direction,  which is clearly due to channel modes of MRI (\citealt{Balbus_hawley1992, Balbus_hawley1998}).  During saturation, both mean fields and EMFs show a coherent vertical structure which changes signs in time with a definite period.  We find magnetic field components $B_y$ and EMF $\bar{\mathcal{E}}_x$ show a very  coherent spatio-temporal variation with a time period of $\approx 9$ orbital period ($2\pi/\Omega$), similar to the earlier studies of MRI dynamo (\citealt{Brandenburg1995, Davis2010, Gressel2010, Gressel2015, Ryan2017}).  This periodicity is semi-transparent in the butterfly diagram of $\bar{B}_x$, while 
this is hardly apparent for $\bar{\mathcal{E}}_y$. 
However, we  note that periodicities exist in all components of mean fields and EMFs as 
will become clear below (see Fig. \ref{fig:psd}).

\subsection{Evolution of kinetic and current helicities}
 \label{sect:kin_curr_helicity}
\begin{figure}
    \includegraphics[scale=0.58]{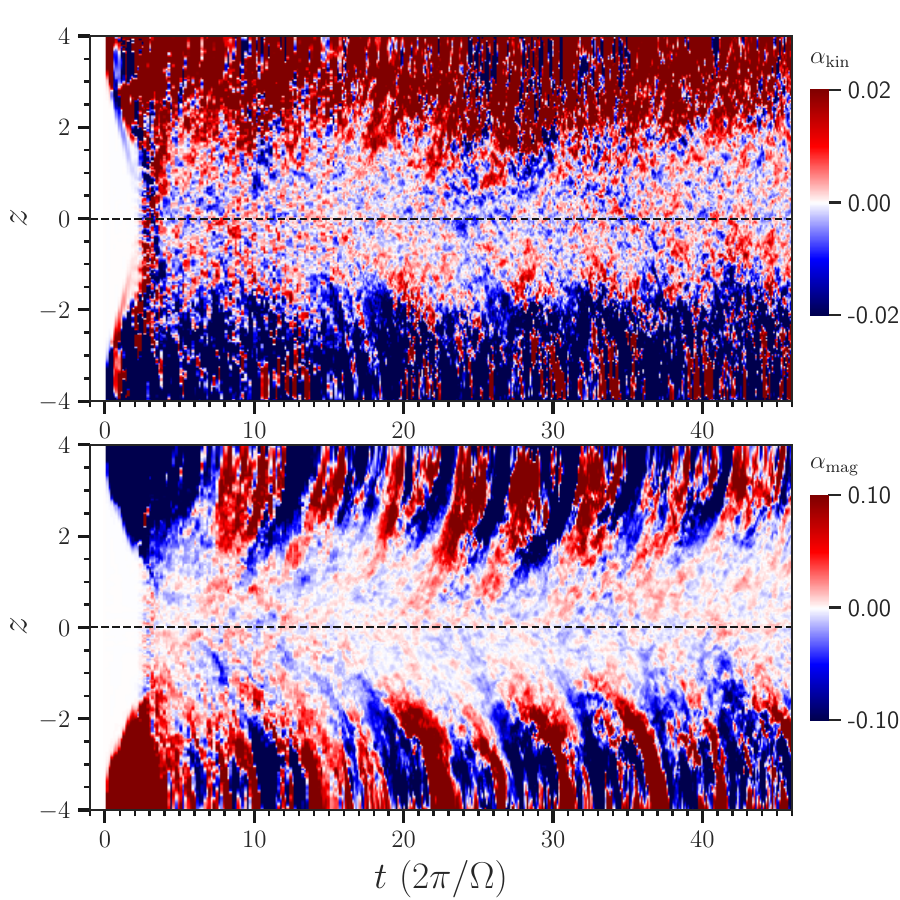}
 \caption{Spatio-temporal variation of $\alpha^{\rm dyn}_{\rm kin} (z,t)$ and $\alpha^{\rm dyn}_{\rm mag} (z,t)$ assuming $\tau_c=\Omega^{-1}$. Both the helicities are small close to the mid-plane, and become larger at larger heights. The $\alpha_{\rm mag}$ flips sign with a time period $\approx 5$ orbital period ($2\pi/\Omega$), roughly half the dynamo period, while $\alpha_{\rm kin}$ does not show any periodicity.  }
   \label{fig:alpha_kin_mag}
 \end{figure}
 
 The generation of large-scale magnetic fields by a dynamo action is often associated with helicity in the fluid velocity field. Assuming isotropic homogeneous turbulence, \cite{Krause_Raedler1980} suggested a kinetic  $\alpha$-effect defined by
 \be
  \alpha_{\rm kin} = -\frac{\tau_c}{3} K_{\rm hel} =- \frac{\tau_c}{3} \  \overline{v^{\prime}.\nabla \times v^{\prime}}  
 \label{eq:alpha_kin}
 \ee 
 responsible for magnetic field generation; where $\tau_c$ is the correlation time, and $K_{\rm hel} = \overline{v^{\prime}.\nabla \times v^{\prime}}$ is the kinetic  helicity. It is suggested  that $\alpha_{\rm kin}$ accounting for the effects of the helical velocity field, takes the role of driver, while $\alpha_{\rm mag}$  (\citealt{Pouquet1976}) defined by  
\be 
 \alpha^{\rm dyn}_{\rm mag} (z,t)=  \frac{\tau_c}{3} C_{\rm hel} = \frac{\tau_c}{3} \overline{v_{A}^{\prime}.\nabla \times v_{A}^{\prime}}  ,
\label{eq:alpha_mag}
\ee 
is the non-linear response arising  due to the Lorentz force feedback, gradually increasing and ultimately quenching the kinetic-$\alpha$ (\citealt{Blackman_Brandenburg2002, Sub2002}). Here, $v^{\prime}_A=\sqrt{B^{\prime 2}/\rho}$ is the Alf\`ven velocity and $C_{\rm hel} = \overline{v_{A}^{\prime}.\nabla \times v_{A}^{\prime}}$ is the current helicity. Ideally, the effective $\alpha-$ effect, responsible for poloidal field generation, is supposed to be $\alpha_{\rm dyn}=\alpha_{\rm kin} + \alpha_{\rm mag}$.

 Fig. \ref{fig:alpha_kin_mag} shows the spatio-temporal variation of $\alpha_{\rm kin}$ and $\alpha_{\rm mag}$. We assume correlation time $\tau_c$ to be same for both $\alpha$-s and  $\tau_c= \Omega^{-1}$. The $\alpha_{\rm mag}$ changes sign with a time-period $\approx 5$ orbital period ($2\pi/\Omega$), roughly half of the dynamo period, with which the mean fields and EMFs change sign, while $\alpha_{\rm kin}$ does not show any explicit periodicity.  We will postpone  a  detailed discussion on the periodicity of helicities to section \ref{sect:PSD} where we discuss periodicities associated with all important variables.

 \subsection{Co-existence of small and large scale dynamos}

 Both kinetic  and magnetic-$\alpha$-s are small close to the mid-plane as shown in Fig. \ref{fig:alpha_kin_mag}, while this is not true of the random kinetic and magnetic energies (  e.g. see Fig. \ref{fig:turb_pump} where we illustrate vertical profiles of rms value of   random fluid velocity and Alfven velocity ). 
 At the same time, the amplitudes of the helicities
 increase away from the mid-plane. These features suggest 
 that both small-scale dynamo (when magnetic field grows because of the random stretching and twisting of the magnetic fields due to turbulent fluid motion) and large-scale dynamo (when magnetic field grows under the action of helicities) co-exist in MRI-driven dynamo (\citealt{Blackman_tan2004, Gressel2010}). 
 The MRI-driven small-scale dynamo dominates magnetic field generation 
 close to the disc mid-plane where stratification is unimportant and helicities are small. 
 In contrast, at larger heights where stratification becomes important, and helicities are large, a helicity-driven large-scale dynamo governs the magnetic field generation (\citealt{Dhang2019, Dhang2020}). 
 However, it is to be noted that $\alpha_{\rm mag}$ is larger than $\alpha_{\rm kin}$ by one order of magnitude, and hence it is very likely that the effective-$\alpha$ will be predominantly due to $\alpha_{\rm mag}$.

\subsection{Power spectra of mean fields, EMFs and helicities}
\label{sect:PSD}
\begin{figure*}
    \centering
    \includegraphics[width=0.9\linewidth]{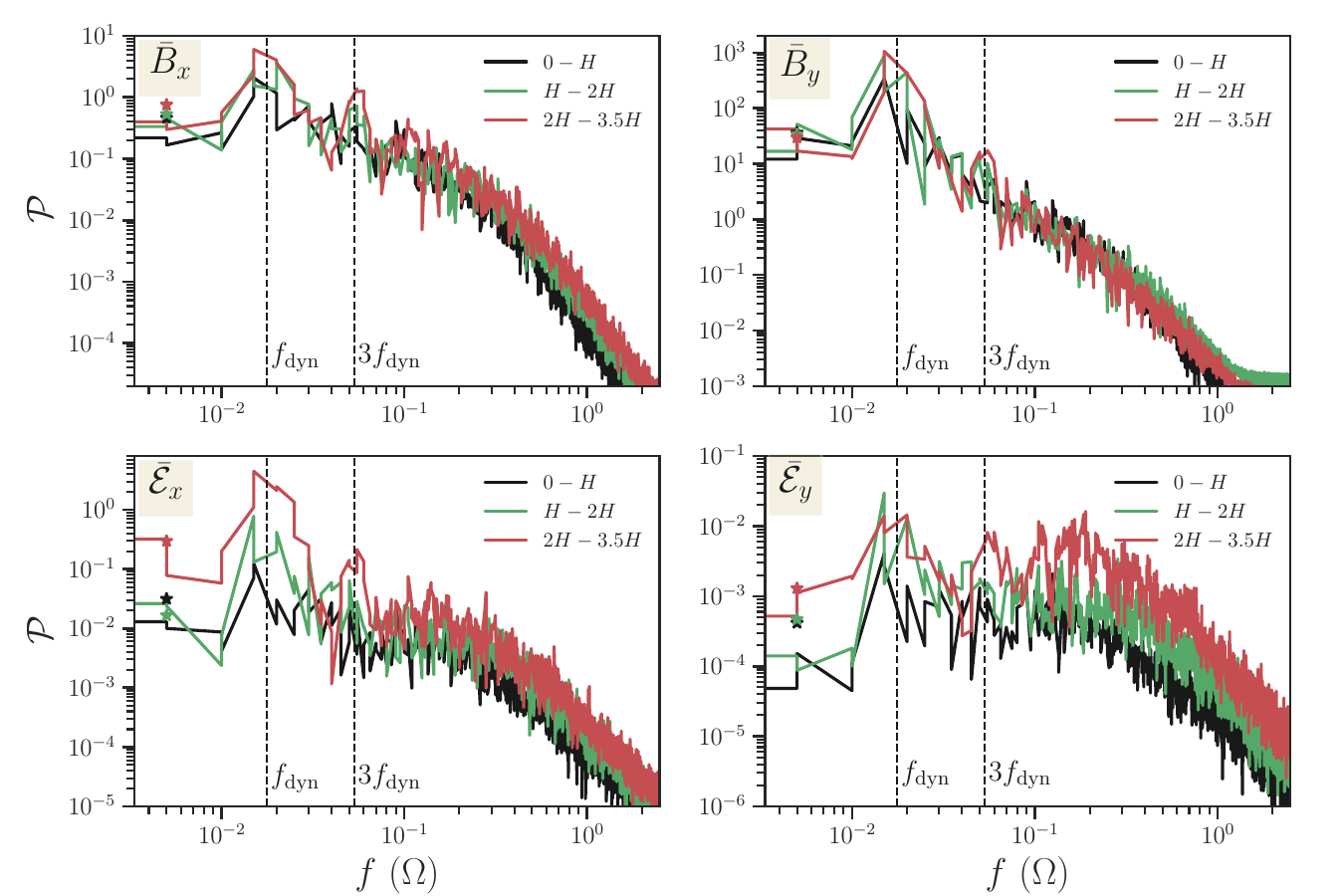}
    \includegraphics[width=0.9\linewidth]{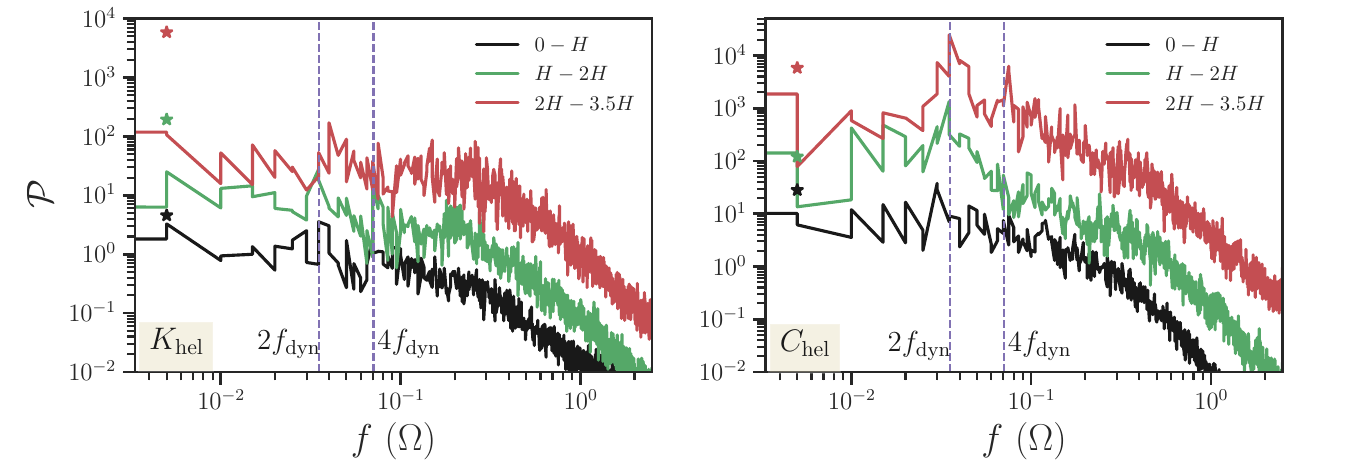}
    \caption{Power spectra of mean fields $\bar{B}_x$, $\bar{B}_y$ (top panels), mean EMFs $\bar{\mathcal{E}}_x$, $\bar{\mathcal{E}}_y$ (middle panels) and helicities $K_{\rm hel}$, $C_{\rm hel}$ (bottom panels). Spatial averages are done over different heights: $z=0-H$ (black lines), $z=H-2H$ (green lines), and $z=2H-3.5H$ (red lines). The zeroth frequency values are denoted by `asterisks'. Vertical dashed lines denote the dynamo frequency $f_{\rm dyn}=0.017$ and its multiples. }
    \label{fig:psd}
\end{figure*}

The butterfly diagrams shown in the previous sections depict the apparent periodicities of mean fields, EMFs and helicities. We look at the power spectrum defined by
\be
\mathcal{P}_{q}(f) = \frac{1}{z_2-z_1}\int_{z_1}^{z_2} dz \  \left |\int \bar{q} (z,t) e^{i f t} dt \right|^2
\ee 
 where $\bar{q} (z,t)$ is any generic quantity to investigate the periodicities in greater detail. Here, the spatial average is done over different heights, namely $z=0-H$, $z=H-2H$ and $z=2H-3.5H$ to study the variation of periodicities at different scale heights.

 Fig. \ref{fig:psd} shows the power spectra of mean fields $\bar{B}_x$, $\bar{B}_y$ (top panels), mean EMFs $\bar{\mathcal{E}}_x$, $\bar{\mathcal{E}}_y$ (middle panels) and helicities $K_{\rm hel}$, $C_{\rm hel}$ (bottom panels). It is noticeable that power spectra for mean fields and spectra peaks at the primary frequency $f_{\rm dyn}=0.017$ $\Omega$ (equivalent to $\approx 9$ orbital period), which was also visible in the butterfly diagrams. In addition to the primary frequency, the power spectra also show the presence of higher harmonics (at $3 f_{\rm dyn}$), which went unnoticed in the earlier works of MRI dynamo. Similarly, power spectra of current helicity $C_{\rm hel}$ also show the presence of higher harmonics (at $4 \ f_{\rm dyn}$) in addition to the primary frequency at $2 f_{\rm dyn}$. We also note that dynamo frequency remains almost constant at different heights. However, kinetic helicity does not show any periodicity. Presence of a strong time variation in $\alpha_{\rm mag}$ and its dominance over $\alpha_{\rm kin}$ necessarily leads to the expectation that turbulent dynamo coefficients ($\alpha-$ coefficients) should harbour a time-dependent part ($\alpha^{1}_{ij}$) along with the traditional time-independent part ($\alpha^{0}_{ij}$) as discussed in section \ref{sect:IROS}.

\section{Results: Dynamo coefficients from IROS}
\label{sect:IROS_result}
We obtained mean fields ($\bar{B}_x$, $\bar{B}_y$), EMFs ($\bar{\mathcal{E}}_x$, $\bar{\mathcal{E}}_y$) from the shearing-box simulation and use a modified version of IROS method (see section \ref{sect:IROS}) to calculate time-independent and time-dependent turbulent dynamo coefficients characterizing the MRI dynamo. However, we find the $x-y$-averaging cannot remove all the signatures of the small-scale dynamo. The small-scale dynamo is expected to have a shorter correlation time of order few $\Omega^{-1}$ and contribute
noise at the higher frequency end compared to the large-scale dynamo.
Therefore, we further smooth the mean fields and EMFs using a low-pass Fourier filter and remove contributions from the frequencies $f>f_c$. We consider three cases: (i) $f_c=0.05 \ \Omega$ ($\approx 3 f_{\rm dyn}$), (ii) $f_c=0.12 \ \Omega $ ($\approx 6 f_{\rm dyn}$) and (iii) $f_c \to \infty$ (unfiltered) to assess the effects of the small-scale dynamo on the dynamo coefficient extraction.

\subsection{Time independent dynamo coefficients}
\begin{figure*}
    \centering
    \includegraphics[width=1.0\linewidth]{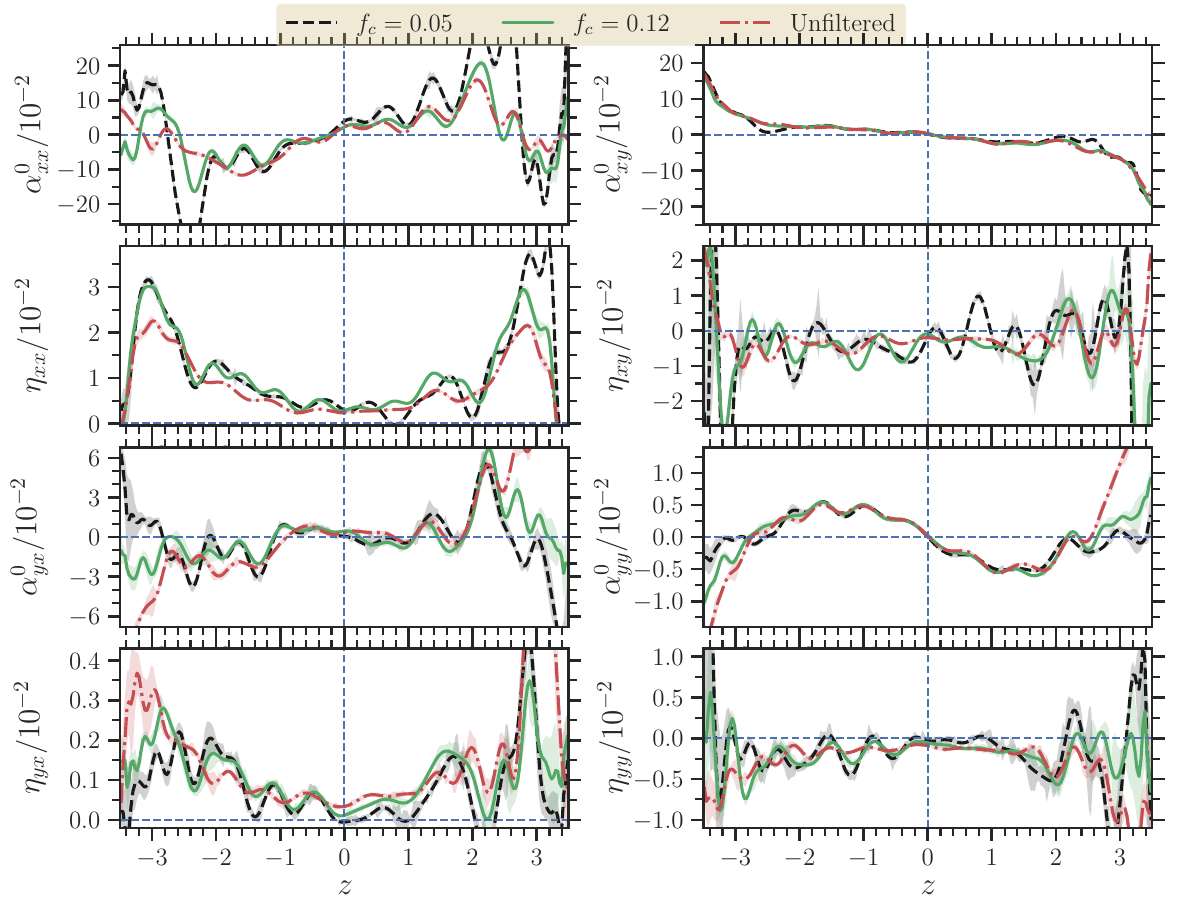}
    \caption{Vertical profiles of time-independent turbulent dynamo coefficients ($\alpha^{0}_{ij}$, $\eta^{0}_{ij}$) in MRI simulation calculated using IROS method.  A low-pass Fourier filter with a cut-off frequency $f_c$ removes the contribution from the small-scale dynamo. We used two values of $f_c$: $f_c=0.05 \ \Omega$ and $f_c=0.12 \ \Omega$. The results are compared to the case when IROS is applied to the unfiltered data obtained from DNS. Shaded regions associated with each line in the plot represent $\pm 1 \sigma$ statistical error on calculated coefficient as described in section \ref{sect:IROS}. }
    \label{fig:coeffs_0}
\end{figure*}

\begin{figure}
    \centering
    \includegraphics[scale=0.5]{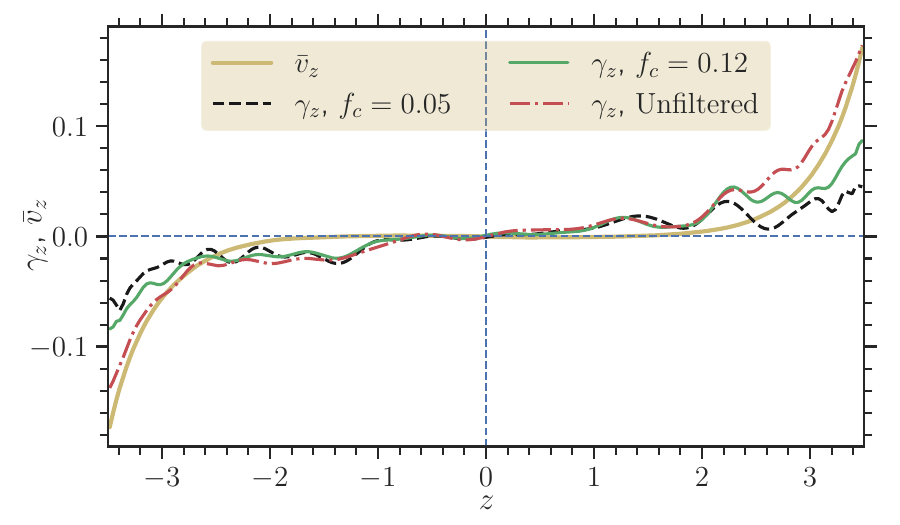}
    \includegraphics[scale=0.5]{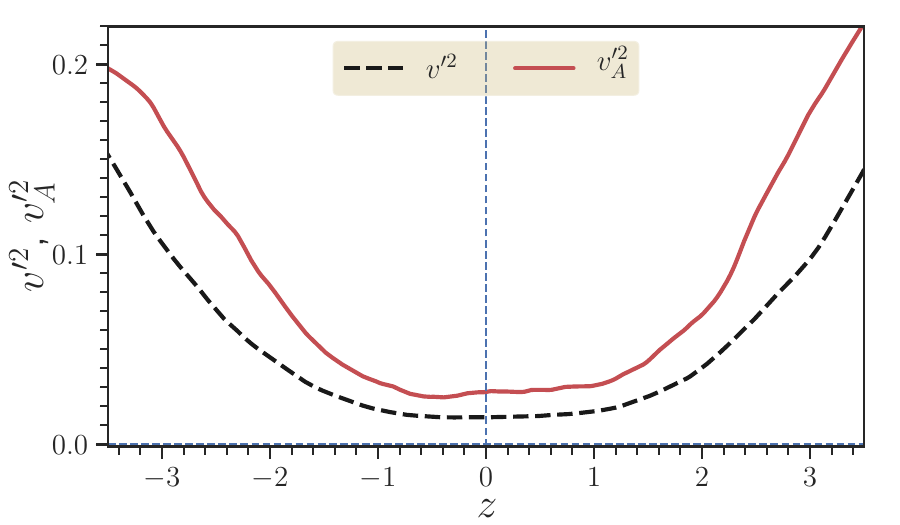}
    \caption{Top panel: profiles of turbulent pumping ($\gamma_z$) and mean vertical outflow ($\bar{v}_z$). They act in the same direction, transporting mean fields vertically outward. Bottom panel: vertical profiles of average fluctuating velocity ($\overline{v^{\prime 2} }$) and fluctuating Alfven speed $\overline{ v^{\prime 2}_A } = \overline{ B^{\prime 2} }/\overline{ \rho } $. Minimal $\tau$ approximation and profiles of $\overline{v^{\prime 2} }$,  $\overline{ v^{\prime 2}_A }$ suggest similar sign of $\gamma_z$ as calculated using IROS.  }
    \label{fig:turb_pump}
\end{figure}

Fig. \ref{fig:coeffs_0} shows the vertical profiles of time-independent dynamo coefficients $\alpha^{0}_{ij}$ and $\eta_{ij}$ for different values of $f_c$.  Four panels at the top illustrate the vertical profiles of coefficients ($\alpha^{0}_{xx}, \ \alpha^{0}_{xy}, \  \eta_{xx}, \ \eta_{xy}$) associated with the x-component of EMF $\bar{\mathcal{E}}_x$, while four panels at the bottom show profiles of those ($\alpha^{0}_{yx}, \ \alpha^{0}_{yy}, \  \eta_{yx}, \ \eta_{yy}$) associated with the y-component of EMF $\bar{\mathcal{E}}_y$.

The  `coefficient of most interest' out of the calculated ones is $\alpha^{0}_{yy}$, which plays a vital role in producing the poloidal field (here $\bar{B}_x$) out of the toroidal field ($\bar{B}_y$) (also see section \ref{sect:mag_energy_contribution}) via an $\alpha$-effect, associated with the helicities (see section \ref{sect:kin_curr_helicity} ). 
The coefficient $\alpha^{0}_{yy}$ shows an anti-symmetric behaviour about the $z=0$ plane, with a negative (positive) sign  in the upper (lower)-half plane (for $|z|<2$). For $|z|>2$, the sign of $\alpha^{0}_{yy}$ tends to be positive (negative) in the upper (lower)-half plane. Earlier studies of MRI dynamo in local (\citealt{Brandenburg2008, Gressel2010, Gressel2015}) and global (\citealt{Dhang2020}) frameworks also found a similar trend in $\alpha^{0}_{yy}$. However, it is to be noted that our study suggests a stronger negative $\alpha^{0}_{yy}$ in the upper-half plane compared to that in the earlier studies. The negative sign in the upper half plane is attributed to the buoyant rise of magnetic flux tubes under the combined action of magnetic buoyancy and shear (\citealt{Brandenburg1998, Brandenburg2005}; see also \citet{Devika+2023}).
\cite{Brandenburg1998} also suggested that negative $\alpha_{yy}$ is responsible for the upward propagation direction of dynamo waves seen in the butterfly diagrams of MRI-driven dynamo simulations (e.g. see Fig. \ref{fig:b_emf_mean}). 
Another different way of looking at the origin of the effective $\alpha$ is to link it to the helicity flux as envisaged by \citealt{Vishniac2015} and \citealt{Gopalakrishnan2023}. We discuss this possibility in section \ref{sect:DCalpha}. 

The off-diagonal terms of the $\alpha$-coefficients are related to turbulent pumping. This effect is responsible for transporting large-scale magnetic fields from the turbulent region to the laminar region. We found $\alpha^{0}_{xy}$ and $\alpha^{0}_{yx}$ to be antisymmetric and $\alpha^{0}_{xy} > \alpha^{0}_{yx}$ unlike the previous studies (\citealt{Brandenburg2008, Gressel2015}) which found $\alpha^{0}_{yx} \approx \alpha^{0}_{xy}$. This resulted in a strong turbulent pumping $\gamma_z = (\alpha^{0}_{yx} - \alpha^{0}_{xy})/2$, transporting large-scale magnetic fields from the disc to the corona as shown in the top panel of Fig. \ref{fig:turb_pump}. We also compare the relative importance of turbulent pumping ($\gamma_z$) and wind ($\bar{v}_z$) in advecting the magnetic field upward (in the upper half-plane) at different heights. Vertical profiles of $\gamma_z$ and $\bar{v}_z$ in the top panel of Fig. \ref{fig:turb_pump} shows that at low heights ($|z|<2.5$), turbulent pumping is the dominant effect over the wind, while the effects of wind become comparable or larger than the pumping term at large scale-heights (see also Fig. \ref{fig:bx2_by2}).

The theory of isotropic kinematically forced turbulence predicts that $\gamma_z$ is supposed to be in the direction of negative gradient of turbulent intensity ($v^{\prime 2}$) (\citealt{Krause_Raedler1980}), that is, in the negative z-direction (in the upper-half plane) in our simulation. This is opposite to what has been found in Fig. \ref{fig:turb_pump}. However, it is to be noted that MRI turbulence in a stratified medium is neither isotropic nor homogenous. Minimal $\tau$-approximation (MTA) suggests that in a stratification and rotation-induced anisotropic turbulent medium, which includes the quasi-linear back reaction due
to Lorentz forces,
\be
\label{eq:gamma_z_mta}
\gamma^{\rm MTA}_z = -\frac{1}{6} \tau \nabla_z ( \overline{v^{\prime 2}} - \overline{B^{\prime ^2}} ) - \frac{1}{6} \tau^2 \bf \Omega \hat{z} \times \nabla_z (\overline{v^{\prime 2}} + \overline{B^{\prime ^2}}),
\ee  
where $\tau$ is the correlation time and it is assumed that $\rho=1$ (see equation (10.59) in \citealt{Brandenburg2005}). The last term in equation \ref{eq:gamma_z_mta} vanishes because all the variables are functions of $z$ alone. Therefore, equation  \ref{eq:gamma_z_mta} and the bottom panel of Fig. \ref{fig:turb_pump} illustrating the vertical profiles of $\overline{v^{\prime 2}}$ and $\overline{v_A^{\prime 2}}$  imply that sign of turbulent pumping obtained from MTA supports that obtained from extracted dynamo coefficients.

We found turbulent diffusion tensor $\eta_{ij}$ to be  anisotropic with $\eta_{xx} > \eta_{yy}$ and having a significant contribution from the off-diagonal components $\eta_{xy}$ and $\eta_{yx}$. Different values of diagonal components of $\eta_{ij}$ imply that mean field components $\bar{B}_x$ and $\bar{B}_y$ are affected differently by the vertical diffusion (also see section \ref{sect:mag_energy_contribution}). It is worth mentioning that  $\eta_{yy} \approx 0$ for the $f_c=0.05$ case, while it is slightly negative for the other two cases. This is somewhat different from the earlier studies (\citealt{Gressel2010, Gressel2015}), which calculated dynamo coefficients using the TF method and found $\eta_{xx}\approx \eta_{yy} > 0$. Out of the two off-diagonal terms of the diffusion tensor, $\eta_{yx}$ is of particular interest. It is suggested that a negative value of $\eta_{yx}$ can generate poloidal fields  by the shear-current effect (\cite{Squire2016}). However, we find $\eta_{yx}$ to be always positive, nullifying the presence of a shear-current effect in our stratified MRI simulation.

Finally, we discuss the effects  of filtering the time series of mean magnetic fields and EMFs on the extracted dynamo coefficients. Fig. \ref{fig:coeffs_0} illustrates how the dynamo coefficients vary if we filter out the contribution above a cut-off frequency $f_c$ with (i) $f_c=0.05 \ \Omega$ ($\approx 3 f_{\rm dyn}$), (ii) $f_c=0.12 \ \Omega$ ($\approx 6 f_{\rm dyn}$) and (iii) $f_c \to \infty$ (unfiltered). Broadly speaking, while the coefficients ($\alpha^0_{xx},\ \alpha^0_{yx}, \ \eta_{xy}, \ \eta_{yy}$ ) associated with $\bar{B}_x$ and its derivative in the mean-field closure (equation \ref{eq:closure}) show larger variations at higher heights with $f_c$, those ($\alpha^0_{xy},\ \alpha^0_{yy}, \ \eta_{xx}, \ \eta_{yx}$) associated with $\bar{B}_y$ and its derivative  are less affected by the filtering process. Especially, $\eta_{yy}$ tends to be more positive with $f_c=0.05$, which is more desirable. To summarize, filtering out the time series of mean magnetic fields and EMFs helps to remove the signature of the small-scale dynamo and to obtain noise free coefficients.  

\subsection{Time dependent dynamo coefficients}

\begin{figure*}
    \centering
    \includegraphics[width=1.0\linewidth]{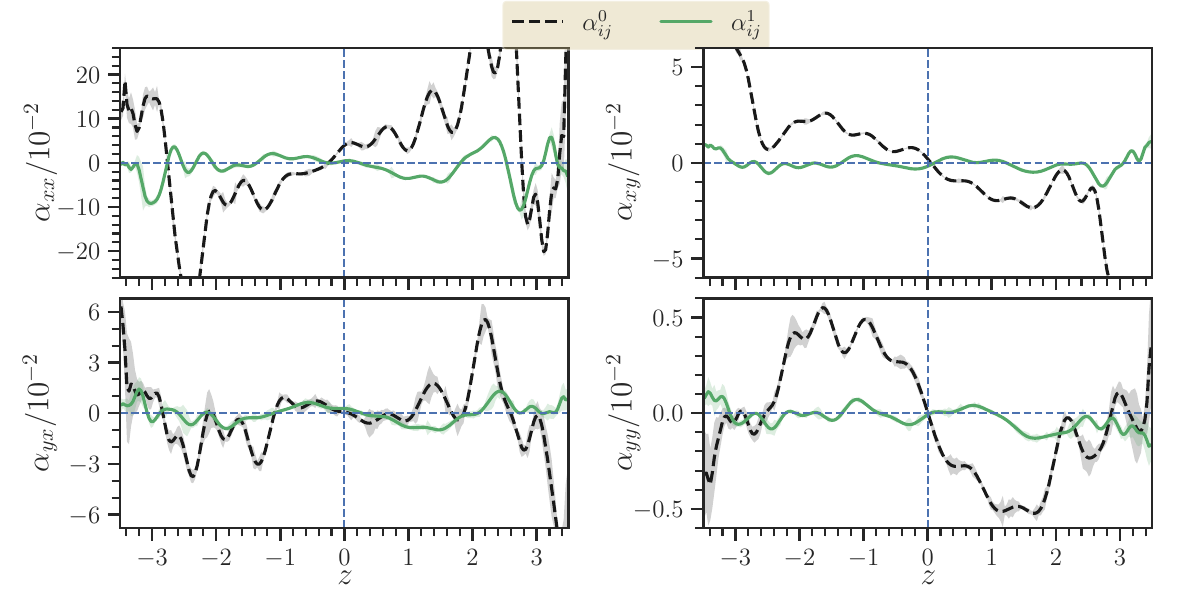}
    \caption{Vertical profiles of time-dependent turbulent dynamo coefficients ($\alpha^{1}_{ij}$) in MRI simulation calculated using IROS method for $f_c=0.05$. Shaded regions associated with each line in the plot represent $\pm 1 \sigma$ statistical error on calculated coefficient as described in section \ref{sect:IROS}.  The time-dependent $\alpha-$s are compared to time-independent $\alpha-$ for the same value of cut-off frequency $f_c=0.05$ $\Omega$. Smaller amplitudes of $\alpha^{1}_{ij}$ compared to that of $\alpha^{0}_{ij}$ implies that that the time-independent $\alpha-$s are predominantly governing the dynamo action. }
    \label{fig:coeffs_m}
\end{figure*}

We discussed the time-dependent nature of $\alpha_{\rm mag}$ in the previous sections. Effective $\alpha$-effect is expected to be determined by $\alpha_{\rm mag}$, especially at the larger scale heights where it is of larger amplitude. While $\alpha$ tensor is expected to have the time-dependent part, $\eta$-tensor is supposed to have only the time-independent part, as it only depends on the turbulent intensity (see section \ref{sect:IROS}).
Fig. \ref{fig:coeffs_m} shows the vertical profiles of time-dependent $\alpha$-tensor components for the fiducial $f_c=0.05 \ \Omega$ case. For comparison, we also plot vertical profiles of the time-independent $\alpha-$s in Fig. \ref{fig:coeffs_m}. We find that the coefficients $\alpha_{xx}$ and $\alpha_{yx}$ associated with $\bar{B}_x$ in the mean-field closure (equation \ref{eq:closure}) have stronger time-dependence compared to those coefficients ($\alpha_{xy}$ and $\alpha_{yy}$) associated with $\bar{B}_y$.  Overall, the amplitudes of $\alpha^{1}_{ij}$ are much smaller than the $\alpha^{0}_{ij}$ implying that the time-independent $\alpha-$s are predominantly governing the dynamo action. Additionally, we observed that (not shown in Fig. \ref{fig:coeffs_m}) $\alpha^{1}_{ij}$-s in the fiducial case ($f_c=0.05$) are relatively smaller  than the other two cases ($f_c=0.12$ and Unfiltered).

\subsection{Verification of method}
To verify the reliability of the determined dynamo coefficients we reconstruct the EMFs using the calculated coefficients and run a  1D dynamo model.

\subsubsection{Reconstruction of EMFs}
\begin{figure*}
    \centering
    \includegraphics[width=0.9\linewidth]{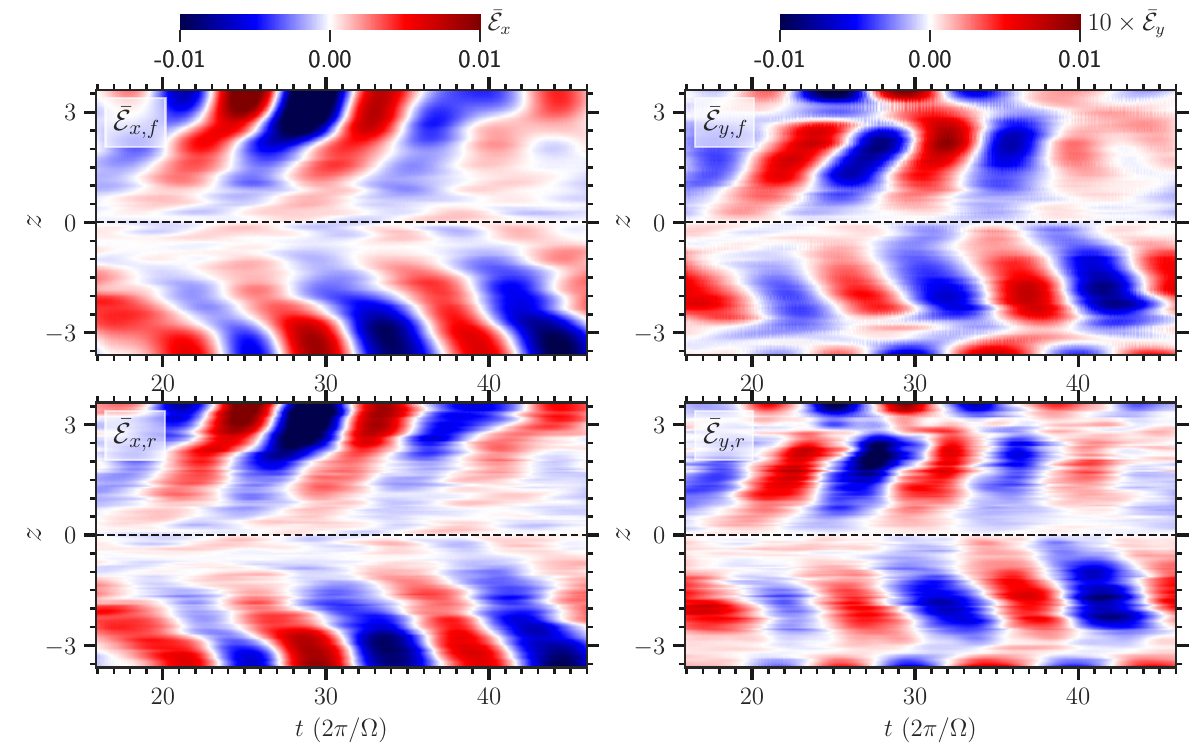}
    \caption{Left panels: Comparison between x-component of EMF $\bar{\mathcal{E}}_{x,f}$ used to determine the turbulent dynamo coefficients,and EMF $\bar{\mathcal{E}}_{x,r}$  reconstructed using the turbulent dynamo coefficients. Right panels: Same as figures in left panels, but for the y-component of EMF. }
    \label{fig:emf_rec}
\end{figure*}

\begin{figure*}
    \centering
    \includegraphics[width=1.0\linewidth]{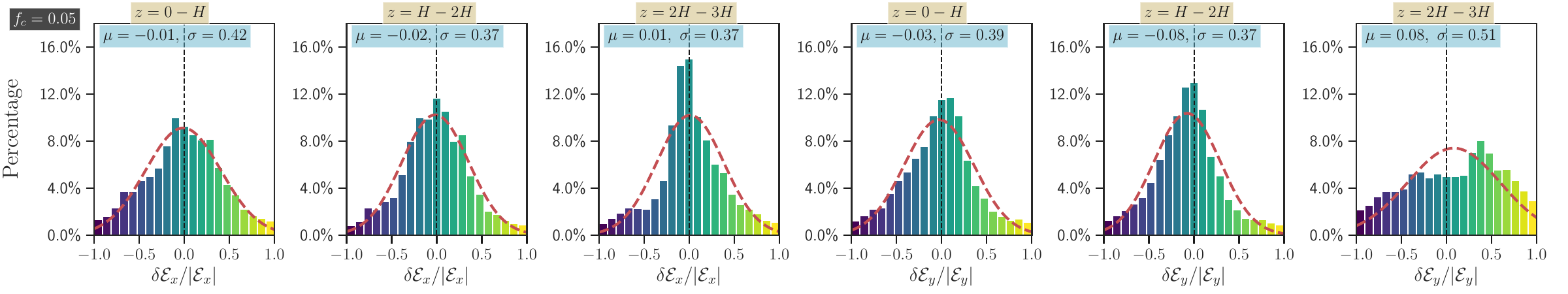}
    \caption{Histograms of the residual EMFs,  $\ \delta  \bar{\mathcal{E}}_{i} =\bar{\mathcal{E}}_{i,f} - \bar{\mathcal{E}}_{i,r} $ calculated within region of different heights  for  $f_c=0.05$ case. We normalise $\delta  \bar{\mathcal{E}}_{i}$ with the absolute values of the corresponding EMFs at the respective points. The Red dashed line shows a normal distribution fitting the histogram.  }
    \label{fig:noise_hist_fc_05}
\end{figure*}

Fig. \ref{fig:emf_rec} shows butterfly diagrams of the EMFs  ($\bar{\mathcal{E}}_{x,f},\ \bar{\mathcal{E}}_{y,f} $) used to determine the turbulent dynamo coefficients and the EMFs ($\bar{\mathcal{E}}_{x,r},\ \bar{\mathcal{E}}_{y,r} $) reconstructed using calculated coefficients and mean fields for $f_c=0.05$. Here it is to be noted that $\bar{\mathcal{E}}_{x,f},\ \bar{\mathcal{E}}_{y,f} $ are the smoothed EMFs obtained by filtering (by using a low-pass filter) EMFs $\bar{\mathcal{E}}_{x},\ \bar{\mathcal{E}}_{y} $ from DNS respectively. We can see a close match between the broad features, such as the dynamo cycle period, in the smoothed and reconstructed EMFs, implying the goodness of fit.

Further,  we investigate the residual of the filtered and reconstructed EMFs, defined by 
\be 
\delta  \bar{\mathcal{E}}_{i} =\bar{\mathcal{E}}_{i,f} - \bar{\mathcal{E}}_{i,r} , \ \ i\in x,y.
\ee 
Fig. \ref{fig:noise_hist_fc_05} shows the histograms of the normalised residuals $\delta \bar{\mathcal{E}}_x/|\bar{\mathcal{E}}_x|$ and $\delta \bar{\mathcal{E}}_y/|\bar{\mathcal{E}}_y|$ calculated within the region of different heights, namely between $0-H$, $H-2H$ and $2H-3H$, for the $f_c=0.05 \ \Omega$ case. All the histograms peak about the region close to zero. However, a  Gaussian fit of the histograms shows that the mean of the distribution always deviates from zero. Additionally, a careful comparison of histograms of  $\delta \bar{\mathcal{E}}_x/|\bar{\mathcal{E}}_x|$ and $\delta \bar{\mathcal{E}}_y/|\bar{\mathcal{E}}_y|$ tells that fit is better for $\bar{\mathcal{E}}_x$ than that for $\bar{\mathcal{E}}_y$, especially at larger scale-heights. Better quality fit for $\bar{\mathcal{E}}_x$ over $\bar{\mathcal{E}}_y$ is expected as  $\bar{\mathcal{E}}_x$ obtained from DNS shows a more regular, coherent space-time variation when compared to $\bar{\mathcal{E}}_y$.

\subsubsection{1D dynamo model}
\label{sect:1d_dynamo}
\begin{figure}
    \centering
    \includegraphics[width=\linewidth]{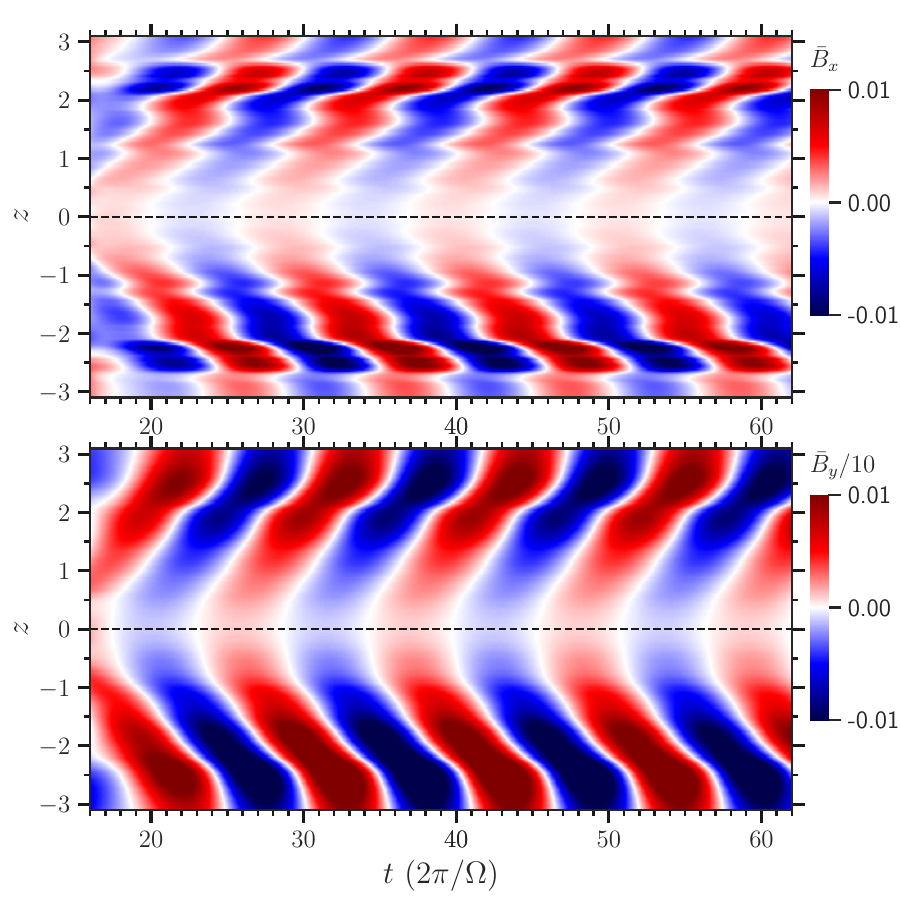}
    \caption{Butterfly diagrams of the mean magnetic fields $\bar{B}_x$ and $\bar{B}_y$ obtained by running 1D dynamo model. Both $\bar{B}_x$ and $\bar{B}_y$ flip sign regularly with a cycle of $\approx 9$ orbital period, similar to that found in shearing box simulations (see Fig. \ref{fig:b_emf_mean}).}
    \label{fig:1d_dynamo}
\end{figure}

We additionally run a 1D dynamo model using the calculated dynamo coefficients and mean velocity field $\bar{v}_z$. 
In particular we solve \eref{eq:mean_field_eq}, or 
in component form 
\begin{align}
\centering
\frac{\partial \bar{B}_x}{\partial t} &=\frac{\partial}{\partial z} \big[ -(\bar{v}_z +\alpha^0_{yx})\bar{B}_x -\alpha^0_{yy}\bar{B}_y + \eta_{yy}\frac{\partial \bar{B}_x}{\partial z} - \eta_{yx} \frac{\partial \bar{B}_y}{\partial z}  \big]\nonumber\\ \nonumber\\
\frac{\partial \bar{B}_y}{\partial t} &=\frac{\partial}{\partial z} \big[ -(\bar{v}_z -\alpha^0_{xy})\bar{B}_y -\alpha^0_{xx}\bar{B}_x + \eta_{xx}\frac{\partial \bar{B}_y}{\partial z} - \eta_{xy} \frac{\partial \bar{B}_x}{\partial z}  \big]\nonumber\\&+q\Omega\,\bar{B}_x.
\label{eq:dynamo}
\end{align}
for $\bar{B}_x$ and $\bar{B}_y$  with
$\alpha^0_{ij}$ and $\eta_{ij}$ obtained using the IROS method. We note  that $\bar{B}_z=0$ as 
a consequence of the zero net flux (ZNF) assumption in our model.
The initial
profiles of $\bar{B}_x$ and $\bar{B}_y$ at  are taken directly 
from the DNS, at time $t=100 \ \Omega^{-1}$ roughly consistent 
with the beginning of the quasi-stationary phase in the DNS. Vertical profile of $\bar{v}_z$ is taken as a constant throughout the evolution and 
is also extracted from the direct simulations by averaging 
it over time throughout the quasi-stationary phase, over which it
roughly stays constant. Additionally, for the profiles of dynamo 
coefficients $\alpha^{0}_{ij}(z)$ and $\eta_{ij}(z)$, we first smooth 
them with a box filter and also cut them off above and below 
three scale heights, and use them in the 1D dynamo model. We do this mainly to avoid the numerical instability 
at boundaries noting that these profiles are sharply flayed 
outside of that range. Note that only the time independent 
parts of the dynamo coefficients are used in the mean field 
equations, since the contributions of $\alpha^1_{ij}$ are
negligible compared to the time-independent part. 

Furthermore, it must be noted that there is a contribution to the 
diffusion from the mesh grids. We do a rough estimation of numerical diffusion as follows $\eta_{0}=v^{\prime}_{\rm rms} \Delta x$, where 
we consider the smallest one among the relevant velocities ($v^{\prime}_{\rm rms},\ c_s, \ v_A$) in the problem. Therefore, we add a correction term $\eta_{0} \approx 10^{-3}$ (with $\Delta x=1/32$ and $v^{\prime}_{\rm rms}=0.1$) to the diagonal components of diffusivity tensor $\eta_{ij}$ to consider the contribution from the mesh to the magnetic field diffusion. This also helps us to stabilize the  1D dynamo solution.

With this setup, we solve the system of equations (equation \ref{eq:dynamo}) with a 
finite difference method over a staggered grid of resolution 
$\Delta z = 1/32$, same as the $z$ resolution of DNS. The outcome 
of this analysis is presented in \fref{fig:1d_dynamo}, where the top and bottom panels  show the butterfly diagrams of $\bar{B}_x$  and $\bar{B}_y$ obtained using the 1D dynamo model respectively. We find both x and y-components of mean fields flip sign regularly with a cycle of $\approx 9$ orbital period, similar to what is found in DNS (see Fig. \ref{fig:b_emf_mean}). Thus, applying calculated coefficients to the 1D dynamo model successfully reproduces broad features of spatiotemporal variations mean magnetic fields.

\subsection{Mean magnetic energy equations}
\label{sect:mag_energy_contribution}
\begin{figure*}
    \centering
    \includegraphics[width=1.0\linewidth]{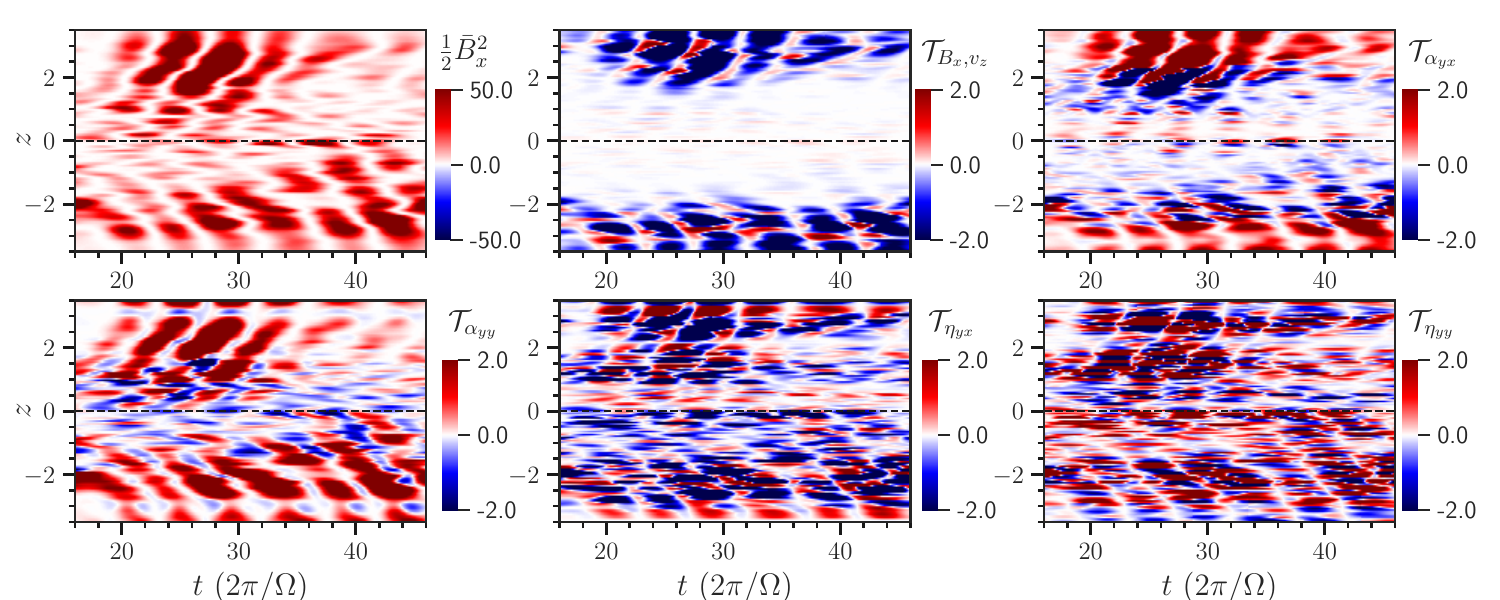}
    \includegraphics[width=1.0\linewidth]{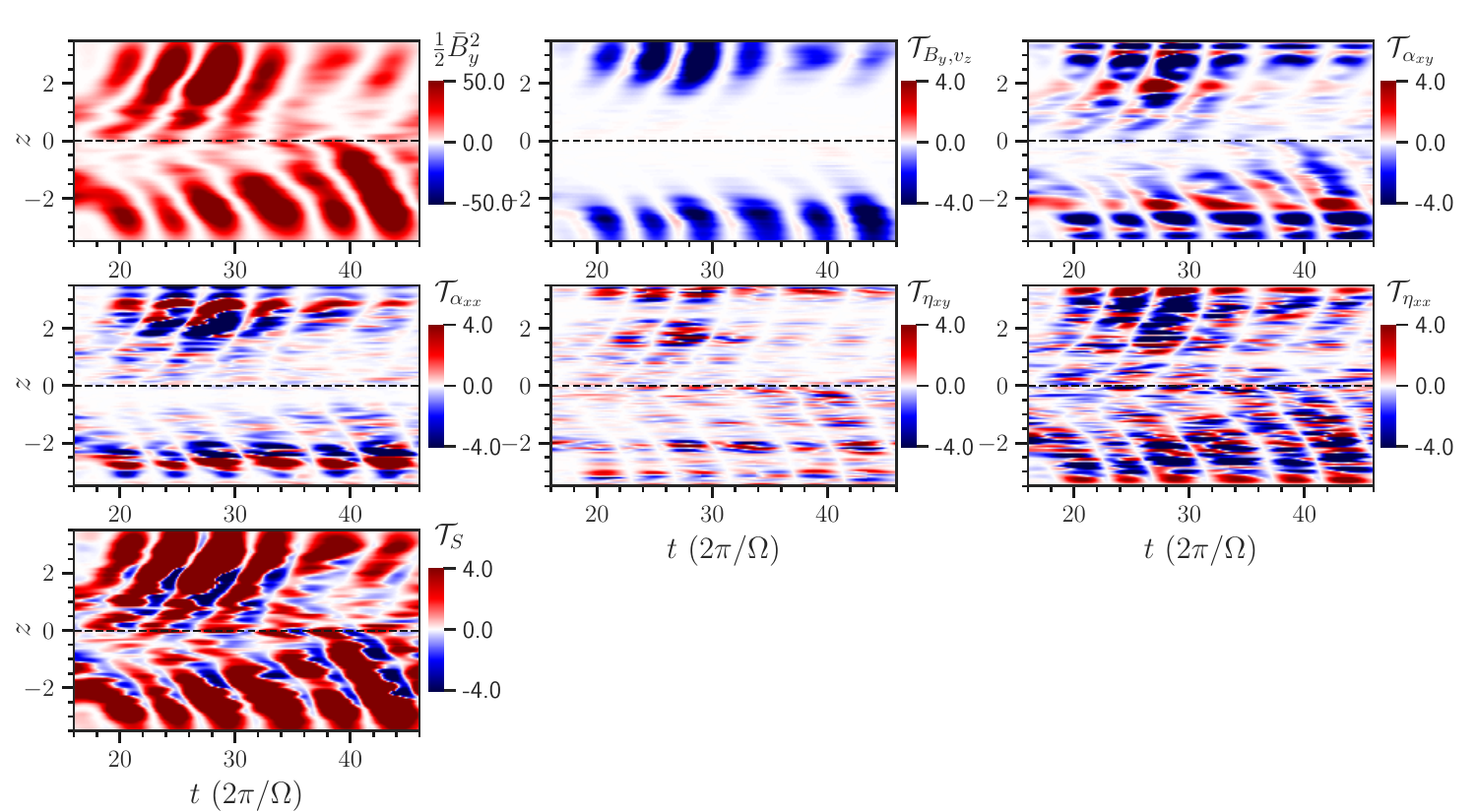}
    \caption{Contributions of different terms involving mean flow ($\bar{v}_z$) and turbulent dynamo coefficients ($\alpha_{ij},\ \eta_{ij}$) to the x-(top six panels) and y-(bottom seven panels) components of  mean magnetic energy evolution equation (equations \ref{eq:bx2} and \ref{eq:by2}). Each term in the poloidal and toroidal magnetic energy equations are multiplied by the factors $10^{6}$ and $10^4$ respectively. Poloidal field ($\bar{B}_x$) generation is primarily attributed to an $\alpha$-effect (the term $\mathcal{T}_{\alpha_{yy}}$), while shear (the term $\mathcal{T}_{\mathcal{S}}$) dominates the toroidal field generation; thus implying and $\alpha-\Omega$ type of dynamo. Winds carry mean fields out of the computational box and contribute largely as the sink term in the mean magnetic energy evolution equation.  }
    \label{fig:bx2_by2}
\end{figure*}
It is challenging to calculate dynamo coefficients uniquely in the presence of both shear and rotation (\citealt{Brandenburg_Radler2008}) as there are  many unknowns (see also discussion in section \ref{sect:diss_closure}). Therefore, it is worth seeing how different terms involving turbulent dynamo coefficients contribute to the mean magnetic energy equation to make physical sense.
The mean magnetic energy evolution equation is obtained by taking the dot product of the mean-field  equation (equation \ref{eq:mean_field_eq}) with the mean magnetic field $\bf{B}$ and given by
\begin{align}
\label{eq:bx2}
  \frac{\partial}{\partial t} \left(\frac{1}{2} \bar{B}^2_x \right) &= \mathcal{T}_{B_x,v_z} + \mathcal{T}_{\alpha_{yx}} + \mathcal{T}_{\alpha_{yy}} + \mathcal{T}_{\eta_{yx}} + \mathcal{T}_{\eta_{yy}},  \\
  \label{eq:by2}
  \frac{\partial}{\partial t} \left(\frac{1}{2} \bar{B}^2_y \right) &= \mathcal{T}_{B_y,v_z} + \mathcal{T}_{\alpha_{xy}} + \mathcal{T}_{\alpha_{xx}} + \mathcal{T}_{\eta_{xy}} + \mathcal{T}_{\eta_{xx}} + \mathcal{T}_{\mathcal{S}} ,
\end{align}
where 
\be
\begin{aligned}
  \mathcal{T}_{B_x,v_z} &= -\frac{1}{2}\bar{B}_x \ \frac{\partial} {\partial z} \left(\bar{v}_z \bar{B}_x \right), \\ 
 \mathcal{T}_{\alpha_{yx}} &= -\frac{1}{2}\bar{B}_x \ \frac{\partial} {\partial z} \left(\alpha_{yx} \bar{B}_x \right), \\
 \mathcal{T}_{\alpha_{yy}} &= -\frac{1}{2}\bar{B}_x \ \frac{\partial} {\partial z} \left(\alpha_{yy} \bar{B}_y \right), \\ 
 \mathcal{T}_{\eta_{yx}} &= -\frac{1}{2}\bar{B}_x \ \frac{\partial} {\partial z} \left(\eta_{yx} \frac{\partial}{\partial z} \bar{B}_y \right),  \\
 \mathcal{T}_{\eta_{yy}} &= \frac{1}{2}\bar{B}_x \ \frac{\partial} {\partial z} \left(\eta_{yy} \frac{\partial}{\partial z} \bar{B}_x \right), \\
 \mathcal{T}_{B_y,v_z} &= -\frac{1}{2}\bar{B}_y \ \frac{\partial} {\partial z} \left(\bar{v}_z \bar{B}_y \right),  \\
 \mathcal{T}_{\alpha_{xy}} &= \frac{1}{2}\bar{B}_y \ \frac{\partial} {\partial z} \left(\alpha_{xy} \bar{B}_y \right),   \\
 \mathcal{T}_{\alpha_{xx}} &= \frac{1}{2}\bar{B}_y \ \frac{\partial} {\partial z} \left(\alpha_{xy} \bar{B}_y \right),  \\
  \mathcal{T}_{\eta_{xy}} &= -\frac{1}{2}\bar{B}_y \ \frac{\partial} {\partial z} \left(\eta_{xy} \frac{\partial}{\partial z} \bar{B}_x \right), \\
   \mathcal{T}_{\eta_{xx}} &= \frac{1}{2}\bar{B}_y \ \frac{\partial} {\partial z} \left(\eta_{xx} \frac{\partial}{\partial z} \bar{B}_y \right), \\
   \mathcal{T}_{\mathcal{S}} &= \frac{1}{2} q \Omega \ \bar{B}_x \bar{B}_y.
\end{aligned}
\ee

Fig. \ref{fig:bx2_by2} shows the space-time plots of different terms involving mean flow ($\bar{v}_z$) and turbulent dynamo coefficients ($\alpha_{ij},\ \eta_{ij}$) in the mean magnetic energy evolution equations. The top six panels in Fig. \ref{fig:bx2_by2} describe the terms in the x-component of the magnetic energy equation (equation \ref{eq:bx2}), while the bottom seven panels illustrate terms in the y-component of the magnetic energy equation ( \ref{eq:by2}) at different heights and times. 

Fig. \ref{fig:bx2_by2} provides a  fairly complicated picture to account for the generation-diffusion scenario of the mean magnetic fields. Broadly speaking, the poloidal field ($\bar{B}_x$) is predominantly generated by an $\alpha$-effect (the term $\mathcal{T}_{\alpha_{yy}}$ in Fig. \ref{fig:bx2_by2}). However, there is a significant contribution from $\alpha_{yx}$ (the term $\mathcal{T}_{\alpha_{yx}}$ in Fig. \ref{fig:bx2_by2}) in generating $\bar{B}_x$ in larger scale-heights. Toroidal field generation is mainly due to the presence of shear, here differential rotation, ($\mathcal{T}_{\mathcal{S}}$ in Fig. \ref{fig:bx2_by2}) which converts poloidal fields to the toroidal fields. However, it is worth noting that there is a minute contribution from the $\alpha_{xx}$, generating a toroidal field out of the poloidal field by an $\alpha$-effect, (as in an $\alpha^2$-$\Omega$ dynamo). The dominance of $\alpha$-effect in generating a poloidal field and that of $\Omega$-effect (shear) in generating a toroidal field imply the presence of an $\alpha-\Omega$ type dynamo in MRI-driven geometrically thin accretion disc. This is similar to what has been found in the study of the dynamo in an MRI-driven geometrically thick accretion disc (\citealt{Dhang2020}), implying universal action of $\alpha-\Omega$-dynamo in MRI-driven accretion flows. 

Generally, it is expected that diagonal components of the diffusion tensor, $\eta_{yy}$ and $\eta_{xx}$ are primarily responsible for the diffusion of $\bar{B}_x$ and $\bar{B}_y$ respectively. However, our simulation finds that winds  carry mean fields out of the computational box and act as a sink in the mean magnetic energy evolution equation, not the $\eta$-s.

\section{Discussion}
\label{sect:discuss}
\subsection{Periodicities in the dynamo cycle}
\label{sect:peirod_dynamo}
Investigations of spatio-temporal variation of different variables in our stratified shearing box simulations  show a diverse range of periodicities. We observed that mean magnetic fields and EMFs oscillate with a primary frequency $f_{\rm dyn}=0.017$ (equivalent to 9 orbital periods), similar to what was found in earlier studies (\citealt{Brandenburg1995, Gammie1996,  Davis2010, Gressel2010, Ryan2017}).   The primary frequency  is determined by the effective dispersion relation of the $\alpha$-$\Omega$ dynamo (e.g. see equation 6.40 in  \cite{Brandenburg2005} and section 3.2 in  \cite{Gressel2015})
with the $\alpha$ dominated by the time-independent (DC) value of $\alpha_{yy}$ .
The plausible origin of 
this  DC value 
of $\alpha_{yy}$ is discussed
below in section ~\ref{sect:DCalpha}.

Additionally, we observed the presence of higher harmonics at $3 f_{\rm dyn}$, which went unnoticed in earlier MRI simulations (see section \ref{sect:PSD}). Unlike the mean fields and EMFs, current helicity shows periodicities at different frequencies $2 f_{\rm dyn}$ and $4 f_{\rm dyn}$, respectively. The presence of the frequencies in the mean EMFs, mean fields, and current helicities can be understood better if we follow the magnetic helicity density evolution equation (e.g.see \cite{BF00, SB06, KR22, Gopalakrishnan2023}),
\be
\frac{1}{2}\frac{\partial h^b}{\partial t} =
-\bar{\mathcal{E}} \cdot \bar{\bf{B}} -\eta_0 \mathcal{C}_{\rm hel} -\frac{1}{2} \nabla\cdot\mathcal{F}_{\mathcal{H}} ,
\label{eq:hel}
\ee 
where $h_b=\la A^{\prime}.B^{\prime} \ra $ is magnetic helicity density, $A^{\prime}$ is the fluctuating vector potential, and   $\mathcal{F}_{\mathcal{H}}$ is the helicity flux. Roughly speaking, magnetic helicity is related to current helicity (and $\alpha_{\rm mag}$, see equation \ref{eq:alpha_mag}) by some length scale and therefore, we can investigate equation \ref{eq:hel} to shed light on the time variation of current helicity.

The component of the EMF along the mean magnetic field
generates mean magnetic and associated current helicities. Now to consider the effect of the DC term  in $\alpha_{yy}$, we assume that this is the dominant term in generating the poloidal field, which is a valid approximation, as we noted in section \ref{sect:mag_energy_contribution} (also see Fig. \ref{fig:bx2_by2}). Then, $\bar{\mathcal{E}}.\bar{\bf{B}} \approx \alpha_{yy}^0\bar{B}_y^2$, which is a source term  in equation \ref{eq:hel}. Now, e.g. for simplicity, if we assume, $\bar{B}_y \sim \sin \ (2\pi f_{\rm dyn} t)$, then magnetic and current helicities which are $\propto \bar{B}^2_y$, will have primary frequency of $2f_{\rm dyn}$. This explains the generation
of magnetic and current helicities at a primary frequency, twice that of $\bar{B}_y$, i.e. 2 $f_{\rm dyn}$. 
This current helicity can now add to the $\alpha$-effect, which combined with the mean field 
in the dynamo equation \ref{eq:mean_field_eq}, can lead to secondary EMF and mean fields components oscillating at $3f_{\rm dyn}$
which in turn sources helicity components at $4f_{\rm dyn}$ and so on. These primary and secondary frequency components, limited by noise, are indeed seen from the analysis of our simulations.

\subsection{Dynamo coefficients, comparison with earlier studies}
Earlier studies calculating turbulent dynamo coefficients using the simulation data and the mean field closure (equation \ref{eq:closure}) in the local (\citealt{Brandenburg1995, Brandenburg2008, Gressel2010, Gressel2015, Shi2016}) and global (\citealt{Flock2012, Hogg2018, Dhang2019, Dhang2020}) simulations of MRI-driven accretion discs used different methods. Earlier local (\citealt{Brandenburg1995, Davis2010}) and most of the global (\citealt{Flock2012, Hogg2018}) studies calculated only the ``coefficient of interest'' $\alpha_{\phi \phi}$ ($\alpha-$ effect) by neglecting the contributions of other terms in the mean-field closure. Many of the local studies (\citealt{Brandenburg2008, Gressel2010, Gressel2015}) use the linear Test Field (TF) method during the run-time to calculate all the coefficients. A few local (e.g. \citealt{Shi2016, Zier2022, Wissing2022},  the current work) and global (\citealt{Dhang2020}) studies used direct methods to quantify dynamo coefficients. However, it is important to  note that while several authors 
used a linear regression method assuming few constraints on the diffusion coefficients (namely, $\eta_{xx}=\eta_{yy}$),  we use the IROS method without any constraints on the coefficients.

Like most of the earlier local and global studies, we find a negative $\alpha_{yy}$ close to the mid-plane in the upper half-plane. However, direct methods seem to capture negative signs better than TF; which can be realised by comparing $\alpha_{yy}$ profiles in our work (also in \cite{Shi2016}) and in \cite{Gressel2010}. Additionally, we find stronger  turbulent pumping (compared to that in the TF method), transporting large-scale magnetic fields from the disc to the corona, similar to that found in global MRI-dynamo studies (\citealt{Dhang2020}). 



 Additionally, for the first time, we ventured to calculate 
 the time-dependent part of $\alpha_{ij}$ inspired by the periodic behaviour of 
 $\alpha_{\rm mag}$. However, we found that the amplitudes of the time-dependent part of $\alpha-$s ($\alpha^{1}_{ij}$) are much smaller than that of the time-independent $\alpha-$s ($\alpha^{0}_{ij}$). Therefore, we suspect that the time-independent
$\alpha-$s are predominantly governing the dynamo action.

Diffusivity coefficients $\eta_{ij}$ in our work are found to be quite different from that in the earlier local studies (\citealt{Brandenburg2008, Gressel2010, Gressel2015, Shi2016}), with $\eta_{xx} \neq \eta_{yy}$ and $\eta_{yy}\approx 0$. Several earlier studies (\citealt{Shi2016, Zier2022}) found $\eta_{yx} <0$) in their unstratified and stratified MRI simulations after imposing  a few constraints on the coefficients (e.g. $\eta_{yy}=\eta_{xx}, \eta_{xy}=0$ etc.) and they proposed shear-current effect \citep{Radler1980,RK04,Squire2016} generating poloidal fields in addition to $\alpha$-effect. Recently, \cite{Mondal2023} carried out statistical simulations of MRI in an unstratified ZNF shearing box and found $\eta_{yx}<0$ proposing `rotation- shear-current effect' and the `rotation-shear-vorticity effect'  responsible for generating the radial and vertical magnetic fields, respectively.   However, like some other studies (TF: \citealt{Brandenburg2008, Gressel2010, Gressel2015}), SPH:\citealt{Wissing2022}), we find $\eta_{yx}\geq 0$, unless we impose a constraint on $\eta_{yy}$ being a positive fraction of $\eta_{xx}$. If we assume $\eta_{yy}=f_{\eta} \ \eta_{xx}$ while calculating the coefficients, we find negativity of $\eta_{yx}$ is an increasing function of the factor $f_{\eta}$ (see Fig. \ref{fig:coeffs_f_eta} and Appendix). However, we find that the quality of fit is compromised slightly and  histograms of the residual of filtered (input) and reconstructed  EMFs get broader (with higher standard deviation) with the assumption $\eta_{yy}=f_{\eta} \ \eta_{xx}$. We refer the reader to see Appendix for details.

\subsection{Helicity flux and the DC $\alpha-$effect}
\label{sect:DCalpha}
The coefficient $\alpha^{0}_{yy}$ represents the $\alpha-$effect responsible for poloidal magnetic field generation out of the toroidal field. We found an anti-symmetric profile of $\alpha^{0}_{yy}$ about the disc-midplane similar to the earlier studies (e.g. see \citealt{Brandenburg2008, Gressel2010}). However, it is to be noted that understanding of the physical mechanism determining the vertical profile of $\alpha^0_{yy}$ is incomplete. E.g. \cite{Brandenburg1998} proposed a buoyancy-driven dynamo to explain the negative sign of $\alpha_{yy}$ in the upper half plane. Here, we propose a different way of looking at the origin of $\alpha-$effect by connecting it to a generative helicity flux. 

\begin{figure}
    \centering
    \includegraphics[width=1.0\linewidth]{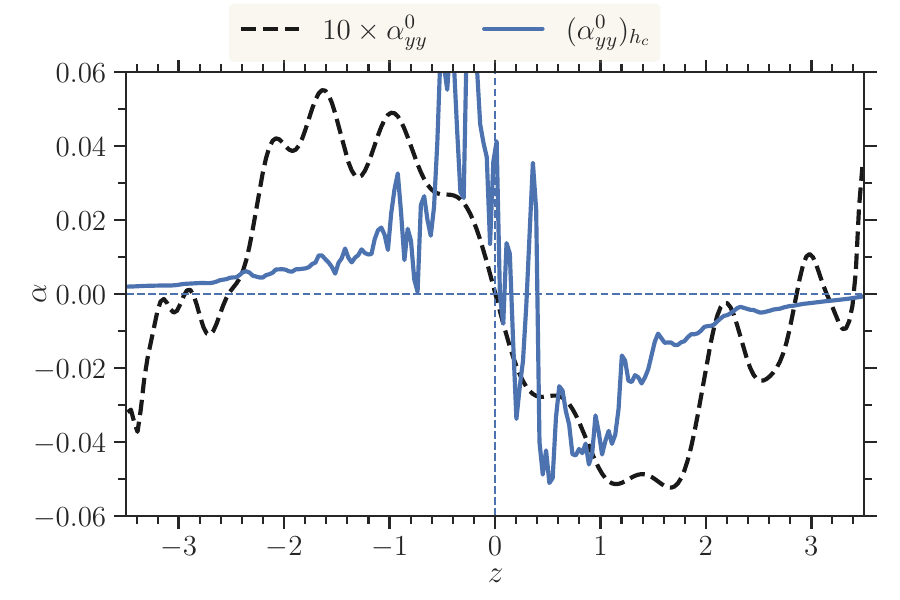}
  \caption{Vertical profiles of  $\alpha^{0}_{yy}$ (for $f_c=0.05$ case) obtained from IROS inversion and $(\alpha^{0}_{yy})_{h_c}$, expected from helicity flux. }
    \label{fig:hel_flux}
\end{figure}

In order to understand the DC value (time-independent) of the $\alpha$-effect, we take the
time average of equation ~\ref{eq:hel}. The term $\partial h_b/\partial t$ averages
to zero, and one gets the well-known constraint \citep{Blackman2016,anv_kan_book}
\begin{equation}
\langle\bar{\mathcal{E}}\cdot\bar{\bf{B}}\rangle=  -\eta_0 \langle\mathcal{C}_{\rm hel}\rangle 
-\frac{1}{2} \nabla\cdot
\langle\mathcal{F}_{\mathcal{H}}\rangle,
\label{eq:steadyEMF}
\end{equation}
where $\langle\rangle$ indicates a time average. This shows that in the absence of helicity fluxes, 
the average EMF parallel to the mean field, responsible
for the generation of poloidal from the toroidal mean field, is resistively (or catastrophically) quenched.
Of the several helicity fluxes discussed in the literature, the
generative helicity fluxes as envisaged in \citet{Vishniac2015} and in \citet{Gopalakrishnan2023}
can source the DC component of $\bar{\mathcal{E}}\cdot\bar{\bf{B}}$ without the preexistence
of any mean field or initial helicities. Using equation (17) of \cite{Gopalakrishnan2023},
with mean vorticity $\Omega (2-q) \hat{z}$
and noting that $\alpha_{yy}\bar{B}_y^2$ dominates  $\bar{\mathcal{E}}\cdot\bar{\bf{B}}$,
we estimate
\be
\begin{split}
\left(\alpha_{yy}^0 \right)_{h_c} \approx -\frac{\Omega\tau^2}{4\langle \bar{B}_y^2\rangle}\Bigg [\left(C_1 v_A'^2  + C_3 v'^2  + C_4 \frac{\lambda^2}{\tau^2}\right) \frac{\partial b'^2}{\partial z} \\  
+ C_2 b'^2 \frac{\partial v'^2}{\partial z}\Bigg],
\end{split}
\label{eq:Genflux}
\ee
where $(C_1,C_2,C_3,C_4) = (7/45,-203/5400,403/8100,-1/6)$
and we have taken $q=3/2$.
Adopting estimates for the correlation time
$\tau\sim \Omega^{-1}$, correlation length $\lambda \sim H/2$
and using the vertical profiles of various physical variables from the simulation,
we calculate the vertical profile of $\left(\alpha_{yy}^0\right)_{h_c}$
due to the generative helicity flux. 
This is shown as a solid line in \fref{fig:hel_flux} and for comparison, we also
show $10 \alpha_{yy}^0$ (for $f_c=0.05$ case) from the IROS inversion.
It is encouraging that the $(\alpha_{yy}^0)_{h_c}$ predicted by the generative helicity flux
is negative in the upper-half plane of the disc and has a qualitatively
similar vertical profile as that determined from IROS inversion. The amplitude, however, is larger,
which perhaps indicates the importance also of the neglected diffusive and advective helicity fluxes which 
act as sink terms in equation \ref{eq:steadyEMF}.

\subsection{Vanishing $\eta_{yy}$, missing information?}
\label{sect:diss_closure}
In section \ref{sect:mag_energy_contribution} we pointed out that 
wind carries away the mean magnetic field and acts as the effective sink of its energy.
However, the poloidal field is also expected to be diffused by $\eta_{yy}$, and a positive $\eta_{yy}$ is required for diffusion. Instead,  we find a vanishingly small (in some regions even negative) $\eta_{yy}$, which leads us to two possible thoughts: either it is impossible to recover $\eta_{yy}$ in the direct methods, or there is incompleteness in the closure we used to retrieve the coefficients. Here, we discuss both possibilities.

It is clear from equation \ref{eq:closure} that the turbulent diffusion coefficients are associated with the currents, which are calculated by taking the $z$-derivative of mean magnetic field components. Calculating derivative makes the currents noisy, especially $J_y$,  as it involves a derivative of $\bar{B}_x$, which is fairly  incoherent over space and time, as can be seen  from the butterfly diagram of $\bar{B}_x$ (Fig. \ref{fig:b_emf_mean}). Additionally, also note that the $y$-component of EMF is also noisy. Thus, the coefficients associated  with $J_y$ and $\bar{\mathcal{E}}_y$ turned out to be error-prone and difficult to calculate. This pattern has been noticed by earlier works (\citealt{Squire2016}), which used direct methods other than IROS, used in the current work. 

In general, mean EMF can be expressed in terms of symmetric, anti-symmetric tensors and mean fields as follows, 

\be 
\label{eq:closure_sym}
 \bar{\mathcal{E}}_i = \tilde{\alpha}_{ij}  \bar{B}_j +  \left( \tilde{\mathbf \gamma} \times \bar{\mathbf B} \right)_{i} - \tilde{\eta}_{ij}  \bar{J}_{j} - \left ( \tilde{\mathbf \delta } \times \bar{\mathbf J} \right)_{i} - \tilde{\kappa}_{ijk} \frac{\partial \bar{B}_{j}}{\partial x_k},  
\ee 
where we neglect the higher than first-order spatial  derivatives and time derivatives of mean fields (\citealt{Radler1980, Brandenburg2005, Schrinner2007, Simard2016}).  The coefficients $\tilde{\mathbf \alpha}$ and $\tilde{\mathbf \gamma}$ represent the symmetric and anti-symmetric parts of $\alpha_{ij}$ tensor in equation \ref{eq:mean_field_eq}. The coefficients $\tilde{\alpha}_{ij}$ is related to $\alpha_{xx}$ and $\alpha_{yy}$, while $\tilde{\gamma}$ represents turbulent pumping. The term $\tilde{\mathbf \eta}$ is a rank-two tensor representing diffusivity. The coefficient $\tilde{\mathbf \delta}$ is interpreted as  a magnetic field generating term \citep{Radler1980}. The $\tilde{\mathbf \kappa}$-term is a third-rank tensor having a complicated influence on mean fields. 

If we define mean fields and EMFs as the $x-y$-averaged quantities, then mean field closure reduces to equation \ref{eq:basic_closure}. The symmetrised coefficients in equation \ref{eq:closure_sym} and non-symmetrised coefficients in equation \ref{eq:basic_closure} are related as
\be
\label{eq:coeff_sym_unsym}
\begin{aligned}
\tilde{\alpha}_{xx} &= \alpha_{xx}, \\
\tilde{\alpha}_{yy} &= \alpha_{yy}, \\
\tilde{\gamma}_z &= \frac{1}{2}\left(\alpha_{yx} - \alpha_{xy} \right),\\
\tilde{\eta}_{xx} + \tilde{\kappa}_{xyz} &= \eta_{xx},\\
\tilde{\eta}_{yy} - \tilde{\kappa}_{yxz} &= \eta_{yy},\\
\tilde{\delta}_z &= \frac{1}{2}\left(\eta_{xy} - \eta_{yx}\right)+\frac{1}{2}\left(\tilde{\kappa}_{xxz}+\tilde{\kappa}_{yyz}\right).
\end{aligned}
\ee 
Therefore it is evident from equation \ref{eq:coeff_sym_unsym}
that it is impossible to decouple a few coefficients (coefficients in the last three identities) as there are more unknown coefficients than independent variables ($\bar{\mathbf B}, \ \bar{\mathcal{E}}$) and the actual diffusion coefficients ($\tilde{\eta}_{ij}$) might be different from the calculated ones ($\eta_{ij}$).

\section{Summary}
\label{sect:summary}
   We carried out stratified zero net flux (ZNF) shearing-box simulations of MRI  using ideal MHD approximation. We characterised 
the MRI-driven dynamo using the  language  of mean field dynamo  theory.  The turbulent  dynamo coefficients in the mean-field closure are calculated using the mean magnetic fields and EMFs obtained from the shearing box simulation. For this purpose, we used a cleaning (or inversion) 
algorithm, namely IROS, adapted to extract the dynamo coefficients. We verified the reliability of extracted coefficients by reconstructing the EMFs and reproducing the cyclic pattern in mean magnetic fields by running a 1D dynamo model. Here we list  the key findings of our work:
\begin{itemize}
    \item We find mean fields and EMFs oscillate with a primary frequency $f_{\rm dyn}=0.017$ $\Omega$ ($\approx 9$ orbital period). Additionally, they have higher harmonics at $3f_{\rm dyn}$. Current helicity $\alpha_{\rm mag}$ has two frequencies $2f_{\rm dyn}$ and $4f_{\rm dyn}$ respectively. These frequencies can be understood from  mean-field dynamo effective dispersion relation and  helicity density evolution equation, respectively (for details, see section \ref{sect:peirod_dynamo}).
    \item Our unbiased inversion and subsequent analysis show that an $\alpha-$effect ($\alpha_{yy}$) is predominantly responsible for the generation of poloidal field (here $\bar{B}_x$) from the toroidal field ($\bar{B}_y$).  The differential rotation creates a toroidal field from the poloidal field completing the cycle; indicating that an $\alpha-\Omega$-type dynamo is operative in MRI-driven accretion disc.
    \item We find  encouraging evidence that the effective 
    DC $\alpha-$effect can be due to a generative helicity flux (section \ref{sect:DCalpha}). 
    \item We find strong wind ($\bar{v}_z$) and turbulent pumping ($\gamma_z$) carry out mean fields away from the mid-plane. Interestingly, 
    they act as the principal sink terms in the mean 
    magnetic energy evolution equation instead of the turbulent diffusivity terms. 
    \item The unbiased inversion finds an almost vanishing $\eta_{yy}$, while $\eta_{xx}$ and $\eta_{yx}$ are positive. However $\eta_{yx}$ and $\eta_{yy}$ are strongly correlated; if one imposes an arbitrary prior that $\eta_{yy} = f_{\eta} \eta_{xx}$, then one finds 
    for increasing $f_{\eta}$,  an increasingly negative $\eta_{yx}$ which has been interpreted as evidence of shear-current effect for generating poloidal fields (see \aref{sect:appen}).
    \item We point out that defining mean fields by planar averaging can necessarily introduce
    degeneracy in determining all the turbulent dynamo coefficients uniquely. This may have important consequences
    for the physical interpretation of the dynamo coefficients  (see section \ref{sect:diss_closure}). 
    
\end{itemize}

\section*{Acknowledgements}
We thank Prateek Sharma, Oliver Gressel, Dipankar Bhattacharya and Xuening Bai for valuable discussions on numerical set-up, dynamo and IROS. We especially thank Dmitri Udzensky for his careful comments on the manuscript. We also thank the anonymous referee for useful suggestions which improve the clarity of the paper.   All the simulations are run using the Computing facility at IUCAA.

\section*{Data Availability}
The data underlying this article will be shared on 
reasonable request to the corresponding author.

\bibliographystyle{mnras}
\bibliography{bibtex_thesis}

\appendix
\cprotect\section{Dynamo coefficients with constraints on $\eta_{yy}$}
\label{sect:appen}
\begin{figure*}
    \centering
    \includegraphics[width=1.0\linewidth]{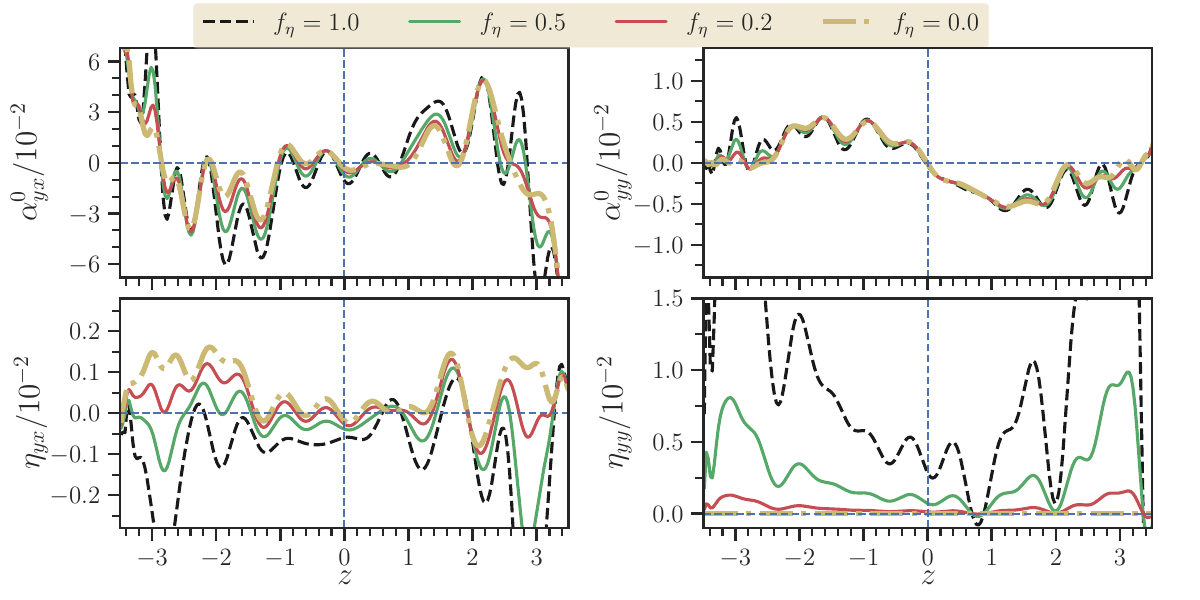}

    \caption{Vertical profiles of the time-independent $\alpha_{ij}$ and $\eta_{ij}$ related to the $\bar{\mathcal{E}}_y$ calculated imposing the constraint $\eta_{yy}=\eta_{xx}$ for $f_c=0.05$ case. As expected, $\alpha_{ij}$ are not affected with the change in $f_{\eta}$, but $\eta_{ij}$s are.  A clear trend has been found; more positive the $\eta_{yy}$ is more negative is $\eta_{yx}$.}
    \label{fig:coeffs_f_eta}
\end{figure*}

\begin{figure*}
    \centering
    \includegraphics[width=1.0\linewidth]{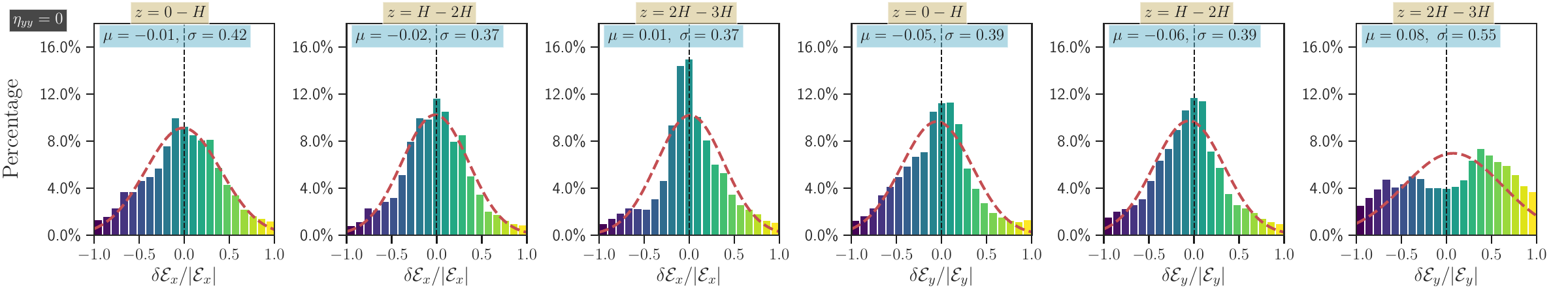}
    \includegraphics[width=1.0\linewidth]{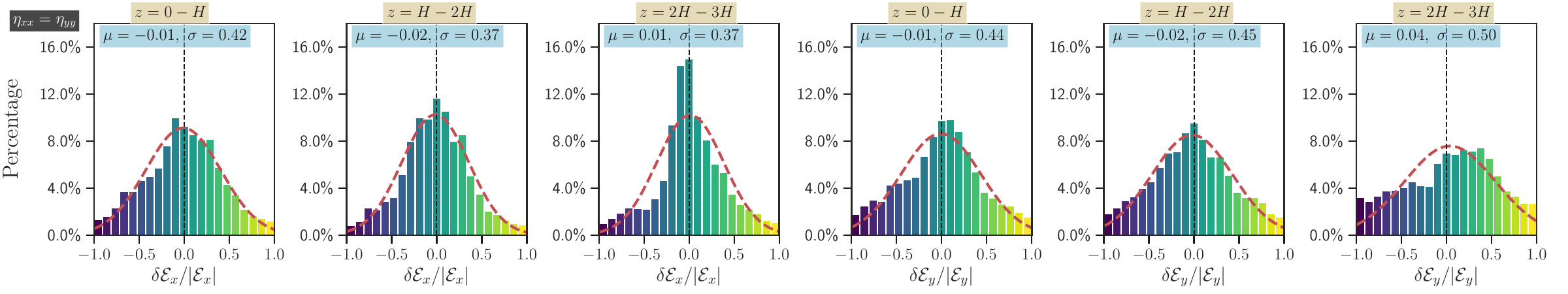}
    \caption{Histograms of the residual EMFs $\delta  \bar{\mathcal{E}}_x$ (top panels) and $\delta  \bar{\mathcal{E}}_y$ (bottom panels)  for $f_{eta}=0$ and $f_{\eta}=1$. We consider the $f_c=0.05$ case. Imposition of constraints on $\eta_{yy}$ compromises the quality of fits, but not significantly because $\alpha_{ij}$ are the main contributors in the fitting of EMFs, not the $\eta_{ij}$.}
    \label{fig:Noise_f_eta}
\end{figure*}
 Diffusivities are challenging to calculate in any direct methods (SVD, linear regression, IROS), as they involve the spatial derivatives of the mean fields. Primarily, we find that $\eta_{yy}$ and $\eta_{xy}$ are noisy as they are related to spatial derivatives of $\bar{B}_x$, which is itself quite noisy (e.g. see butterfly diagram in Fig. \ref{fig:b_emf_mean}. Some earlier studies (\citealt{Squire2016, Shi2016}) put  constraints on calculating $\eta$-s, try to alleviate this issue. E.g. \cite{Shi2016} imposed the constraint that
 $\eta_{yy}=\eta_{xx}$ in the shearing box simulation of MRI and found  a negative $\eta_{yx}$, implying the presence of a shear-current effect.

We have, on the other hand, done an unbiased inversion, as it is not clear if such constraints are actually obeyed by MRI-driven turbulence. Nevertheless, for completeness, we explore here  a more generalized constraint on $\eta_{yy}$, given by  $\eta_{yy}=f_{\eta}\ \eta_{xx}$, and calculate only those coefficients that appear in the mean-field closure for $\bar{\mathcal{E}}_y$, as those related to $\bar{\mathcal{E}}_x$ remain unaffected. Fig. \ref{fig:coeffs_f_eta} shows the vertical profiles of $\alpha^{0}_{yx}$, $\alpha^{0}_{yy}$, $\eta_{yx}$ and $\eta_{yy}$ for different values of $f_{\eta}$ and for $f_c=0.05$. The coefficients $\alpha_{ij}$ remain almost unaffected, while $\eta_{ij}$
 change significantly with change in $f_{\eta}$. There is a clear trend that the more positive the $\eta_{yy}$ (or larger the imposed $f_{\eta}$), 
 the  more negative is $\eta_{yx}$.
 This implies a clear correlation between $\eta_{yx}$ and $\eta_{yy}$. 

 Further, we investigate  the histograms of the residual EMFs $\delta  \bar{\mathcal{E}}_i$ to check the goodness of the fits. The $x-$components of the residual EMFs remain unaffected as expected, while 
 histograms for $\delta \bar{\mathcal{E}}_y$ get slightly broader with the increase in $f_{\eta}$. This implies that the imposition of constraints on $\eta_{yy}$ compromises the quality of fits, but not greatly  
 because $\alpha_{ij}$ are the significant contributors in the fitting of EMFs, not the $\eta_{ij}$.

\label{lastpage}

\end{document}